\documentclass[bibyear]{aa} 

\usepackage{graphicx}
\usepackage{txfonts}
\usepackage{wasysym}
\usepackage{siunitx}
\usepackage{float}
\usepackage{natbib}
\usepackage{amsmath}
\usepackage[colorlinks=true,citecolor=blue,linkcolor=blue]{hyperref}
\usepackage{threeparttable}
\AtBeginDocument{}

\begin{document}

\title{Unveiling the hot interstellar medium in nearby galaxies with eROSITA}

\titlerunning{Hot interstellar medium in nearby galaxies with eROSITA}

   \author{Roman Laktionov\inst{1}
          \and 
          Manami Sasaki\inst{1}
          \and
          Elias Kyritsis\inst{2,3,4}
          \and
          Andreas Zezas\inst{3,4}
          \and
          Steven Hämmerich\inst{1}
          \and
          Fiona Knies\inst{1}
          \and
          J\"orn Wilms\inst{1}
          \and
          Frank Haberl\inst{2}
          \and
          Martin G. H. Krause\inst{5,1}
          \and
          Ashley Coombs\inst{5}
          \and
          Philipp Weber\inst{1}
          \and 
          Aafia Zainab\inst{1}}
          
\authorrunning{Laktionov et al.}

   \institute{Dr. Karl Remeis Observatory, Erlangen Centre for Astroparticle Physics, Friedrich-Alexander-Universität Erlangen-Nürnberg, Sternwartstraße 7, 96049 Bamberg, Germany
              \email{rom.laktionov@fau.de}
              \and
              Max-Planck-Institut f{\"u}r extraterrestrische Physik, Gie{\ss}enbachstra{\ss}e 1, 85748 Garching, Germany 
              \and
              Physics Department \& Institute of Theoretical \& Computational Physics, University of Crete, 71003 Heraklion, Crete, Greece
              \and
              Institute of Astrophysics, Foundation for Research and Technology-Hellas, GR 71110 Heraklion, Greece
              \and
              Centre for Astrophysics Research, University of Hertfordshire, Hatfield AL10 9EU, United Kingdom
             } 
   \date{Received 24 June 2026; accepted 9 October 2026}

  \abstract
   {Diffuse X-ray emission in galaxies arises from both stellar-feedback-heated gas and unresolved compact sources. Previous studies mainly focused on the integrated X-ray luminosity, leaving the thermal component, especially in low-star-formation-rate (SFR) galaxies, largely unexplored.}
   {We aim to characterize the diffuse X-ray emission of nearby galaxies over a wide range of SFRs and to test how the X-ray luminosities $L_\mathrm{X}$ scale with star formation in a blind, volume-limited sample.}
   {We analyzed data of the four eROSITA/\textit{Spectrum-Roentgen-Gamma} all-sky surveys (eRASS:4) for 140 galaxies within 25\,Mpc, drawn from the Updated Nearby Galaxy Catalog, and cross-matched with HECATEv2. Early-type galaxies were treated separately, while the star-forming galaxies were divided by specific SFR (sSFR) into low-, medium-, and high-sSFR subsamples. We extracted spectra of the integrated emission and modeled them with thermal plasma and power-law components. For each subsample, we stacked the spectra and derived total and thermal-component X-ray luminosities. We then fitted relations for the star-forming galaxies.}
   {The stacked spectra show a systematic change with sSFR: high-sSFR galaxies exhibit a pronounced soft thermal component, whereas low-sSFR and early-type galaxies are increasingly dominated by the harder, nonthermal continuum. For the total 0.5--8\,keV luminosity, we obtain $\log L_\mathrm{X} \sim 39.5 + 0.9\,\log \mathrm{SFR}$, with an intrinsic scatter of ${\sim}0.9\,\mathrm{dex}$. For the 0.5--2\,keV luminosity of the thermal component, we find $\log L_\mathrm{X} \sim 38.7 + 1.1\,\log \mathrm{SFR}$, with a smaller intrinsic scatter of ${\sim}0.7\,\mathrm{dex}$.}
   {The X-ray luminosity of the thermal component remains well correlated with the SFR down to ${\sim}10^{-5}\,\mathrm{M_\odot\,yr^{-1}}$ and is consistent with previous relations for star-forming galaxies, while the total $L_X-\mathrm{SFR}$ relation is flatter and more scattered. This result suggests that, at low SFR, the total X-ray output is likely dominated by a small number of unresolved X-ray binaries, whereas the thermal X-ray luminosity remains a robust tracer of stellar-feedback-related X-ray emission.}

   
   \maketitle

\section{Introduction}

Galactic evolution and emission characteristics are strongly affected by the interplay of star formation and stellar feedback in the interstellar medium (ISM). Star formation takes place in cold molecular gas clouds with masses of $10^4 \mbox{--} 10^7\,\mathrm{M_{\astrosun}}$ \citep[][and references therein]{intr1} and behaves differently depending on the galactic environment. In star-forming galaxies, the ISM is heated via shock waves that are produced by stellar winds of massive stars and supernovae. The combined effects of stellar winds and supernova explosions can lead to the formation of superbubbles and giant supershells, which contain hot, ionized plasma with particle densities of $n < \SI{0.01}{cm^{-3}}$ at temperatures of $T {\sim}  10^6 \mbox{--} 10^7\,\mathrm{K}$ (e.g., \citealt{weaver1977,maclow1988, krause2013, krause2014}). The hot thermal plasma  will persist in the galaxy for millions of years and produce diffuse X-ray emission, which again plays a key role for the star-forming activity in the galaxy. Further processes that greatly affect star formation and galactic evolution are tidal interactions and mergers with other galaxies. The accreted matter can initiate intense bursts of star formation over short time periods, so-called starbursts. Starburst regions exhibit a high density of massive stars, which deposit large amounts of energy into the surrounding medium and greatly affect the gas distribution and dynamics of the host galaxy.
By studying the diffuse X-ray emission of nearby galaxies, we can constrain the distribution and spectral properties of the hot, ionized gas and its relation to star-forming activity.

The \textit{Einstein} and \textit{ROSAT} missions have already found in the 1990s that spiral galaxies emit soft X-rays in the 0.1--2.4\,keV band (\citealt{intr2,intr3,intr4}), with the two dominating components being X-ray binaries (XRBs) and hot, thermal plasma.  However, detailed systematic studies of diffuse X-ray emission from galaxies have not been possible due to the lack of spatial resolution in X-rays, which made it hard to separate point sources from diffuse emission. Moreover, emission in the soft X-ray band is highly affected by foreground absorption and contaminated by the X-ray background. For this reason, detailed studies of diffuse X-ray emission have so far been restricted to the brightest and nearest systems, e.g., M83, M101, or NGC 4649 \citep{intr6,intr7,intr8,intr9}. Some studies explore larger samples, e.g., \citet{mineo2}, who analyzed 29 star-forming galaxies using \textit{Chandra} data and derived a linear relation between the X-ray luminosity of the diffuse gas $L_\mathrm{X}$ and the galaxy's star formation rate (SFR). However, these samples are limited to relatively high SFRs and are not representative of the typical galaxy population in the Universe. The eROSITA all-sky survey fundamentally changes this situation. With its full-sky coverage and high sensitivity in the 0.2--10\,keV band, eROSITA provides, for the first time, a statistically complete, volume-limited data set that eliminates the strong selection bias toward bright or large systems. After four completed scans (eRASS1 - eRASS4), the survey reaches sufficient depth for a stacking analysis of nearby galaxies, including those for which no pointed \textit{Chandra} or \textit{XMM-Newton} observations exist. The stacking analysis allows us to probe diffuse emission well below the per-galaxy detection threshold. Consequently, we are able to investigate the hot ISM across the full range of SFRs, and to derive an unbiased $L_\mathrm{X}$-$\mathrm{SFR}$ relation for the local galaxy population.

In this study, we analyze the X-ray properties of a volume-limited sample of 140 nearby galaxies from the Updated Nearby Galaxy Catalog \citep{karachentsev}. We first analyze three of the most luminous galaxies in our sample, NGC~5236, NGC~1313, and NGC~1291, as well as an AGN control galaxy outside of our sample, NGC~4594, individually. These galaxies are bright enough to allow spectral modeling, span a wide range of physical conditions, and serve as benchmark cases that illustrate how diffuse X-ray emission appears in single galaxies in eRASS:4. For the fainter galaxies, we studied the integrated X-ray emission in stacked subsamples according to morphology and specific SFR (sSFR; SFR divided by stellar mass). We determined the X-ray luminosities of the galaxies $L_\mathrm{X}$ and derived a correlation with the SFR for the thermal component and the total X-ray emission; including galaxies with SFRs as low as ${\sim}10^{-5}\,\mathrm{M_\odot\,yr^{-1}}$. In Sect.\,\ref{chap2} we present the X-ray data and galaxy sample analyzed in this work. In Sect.\,\ref{chap3} we describe the analysis of the X-ray data. The results are presented in Sect.\,\ref{chap4} and discussed in Sect.\,\ref{chap5}. The findings of this work are summarized in Sect.\,\ref{chap6}.

\section{Data}\label{chap2}

\subsection{Galaxy catalogs}\label{sect_catalogs}

We used the Updated Nearby Galaxy Catalog \citep{karachentsev}, which is an all-sky catalog of 869 nearby galaxies with individual distance estimates out to $d{\sim}\SI{25}{Mpc}$. It contains a wide range of observables, including angular diameters, distance estimates, and morphological classifications, providing the groundwork for a volume-limited study of nearby galaxies. The positions and diameters, given by the 26.5\,mag Holmberg isophote in the \textit{B} band, were used to extract the corresponding eROSITA images and spectra.

We complemented these data with the second release of the Heraklion Extragalactic Catalogue (HECATEv2; \citealt{hecatev2}). HECATEv2 is a value-added catalog of 204733 galaxies that combines multi-wavelength photometry and spectroscopic measurements from major surveys (e.g., SDSS, \textit{2MASS}, and WISE). The catalog provides homogeneous estimates of star-formation rates and stellar masses $M_\ast$, morphological classifications, and structural parameters such as sizes and position angles. The adopted SFRs in HECATEv2 are homogenized from several literature sources, primarily the GALEX+WISE-based measurements of \citealt{leroy2019} (L19), the UV-to-optical spectral energy distribution fits of \citealt{salim2018} (S18), and the infrared/optical calibrations of \citealt{Kouroumpatzakis2023} (K23). Different SFR indicators can introduce systematic variations due to their calibrations and photometric apertures. HECATEv2 accounts for these differences by correcting the measurements to a common distance scale and by rescaling the S18 estimates to the K23 calibration. After homogenization, the S18 and K23 estimates agree with a scatter of $\lesssim 0.3\,\mathrm{dex}$. The L19 and K23 estimates are roughly consistent at $\mathrm{SFR} \gtrsim 0.5\,\mathrm{M_\odot\,yr^{-1}}$, while the scatter increases, mainly due to aperture differences in nearby galaxies, toward lower SFRs. HECATEv2 therefore preferentially adopts the L19 measurements where available.

To combine the two catalogs, we cross-matched the galaxies using the CDS X-Match\footnote{\url{https://cdsxmatch.u-strasbg.fr/}} service with a $10''$ search radius. We adopted this radius to allow for small differences in the cataloged galaxy centers, while keeping it sufficiently small to avoid false identifications.

\subsection{Galaxy sample}\label{select_sample}

We excluded systems lacking SFR or stellar-mass entries in HECATEv2 and galaxies classified as active galactic nuclei (AGN) or low-ionization nuclear emission-line regions (LINERs) in the NASA/IPAC Extragalactic Database (NED; \citealt{ned}). HECATEv2 also provides activity classifications, identifying galaxies as star-forming, AGN, LINER, passive, or composite; but this classification is not available for every galaxy. Therefore we use the HECATEv2 numerical Hubble Type $T$, which designates the morphological classification of a galaxy according to \citet{devaucouleurs1959}, to separate the sample into star-forming/late-type and quiescent/early-type galaxies (ETGs). Galaxies with $T \geq 1$ were assigned to the star-forming sample, while those with $T < 1$ were considered potential ETGs. We then compared this morphology-based separation with the available HECATEv2 activity classifications and the literature. This comparison revealed that seven potential ETGs show evidence for ongoing star formation in the literature, and that NGC~3299 is classified as an AGN in HECATEv2. Therefore we included those galaxies in the star-forming sample and excluded NGC~3299.

This selection resulted in a working sample of 140 galaxies, including 13 ETGs. The ETGs were treated separately, as they exhibit different X-ray emission mechanisms. The late-type galaxies were further grouped into three subsamples according to their specific star formation rate (sSFR; SFR divided by stellar mass $M_\mathrm{\ast}$), which is a measurement of their star-forming intensity. This method separates galaxies based on their relative star-forming activity, rather than absolute SFR which strongly correlates with stellar mass (see right panel in Fig.~\ref{mass_sfr} in Appendix~\ref{appendix_metal}). The sSFR distribution (left panel in Fig.~\ref{mass_sfr}) shows three characteristic regimes: a strong concentration of galaxies below $0.5\cdot10^{-10}\,\mathrm{yr^{-1}}$, an intermediate plateau between $0.5\cdot10^{-10}$ and $2.0\cdot10^{-10}\,\mathrm{yr^{-1}}$, and a high-sSFR tail above $2.0\cdot10^{-10}\,\mathrm{yr^{-1}}$. We therefore adopt these values as the boundaries of the low-, medium-, and high-sSFR galaxy samples. These thresholds are also astrophysically motivated, as they distinguish galaxies lying roughly below, around, and above the star-forming main sequence with sSFRs on the order of $10^{-10}\,\mathrm{yr^{-1}}$ (e.g., \citealt{renzini2015}). The ETGs all lie below the adopted low-sSFR threshold and occupy the lowest end of the sSFR distribution (left panel in Fig.~\ref{mass_sfr}), with a mean sSFR of $3.3\cdot10^{-12}\,\mathrm{yr^{-1}}$. For comparison, the mean sSFR of the low-sSFR subsample equals $2.0\cdot10^{-11}\,\mathrm{yr^{-1}}$. By grouping the galaxies in this way, each subsample represents systems with comparable physical conditions, i.e., thermal gas fraction and stellar feedback strength. In total, we obtain 22 high-sSFR, 48 medium-sSFR, and 57 low-sSFR galaxies.

\subsection{eROSITA}

eROSITA (\citealt{predehl}) is an X-ray telescope developed by a German consortium. It was launched on July 13, 2019 as a part of the Russian-German \textit{Spectrum-Roentgen-Gamma} (\textit{SRG}) mission (\citealt{sunyaev}). eROSITA is highly sensitive at energies between 0.2 and 8\,keV and covers a large field of view (FOV) with a diameter of one degree. The telescope consists of seven telescope modules (TMs), each equipped with a CCD detector, of which all apart from TMs 5 and 7 are equipped with a thin on-chip aluminum optical blocking filter. These two modules are contaminated by optical light from a light leak in the camera housing \citep{predehl}. The telescopes have an average on-axis spatial resolution of $16.1"$ half-energy width at $1.5\,\mathrm{keV}$ (\citealt{predehl,merloni}). It has been agreed to divide the observational data; the western Galactic hemisphere has been assigned to the German consortium, and the eastern hemisphere to the Russian consortium.

\section{X-ray analysis}\label{chap3}

We processed the eROSITA data accessible to the German eROSITA consortium using a data-processing framework, which operates upon the eROSITA Science Analysis Software System (\texttt{eSASS}, version 211214-p4; \citealt{brunner}), following the approach of \citet{haemmerich2026}. Event lists and images were created for each galaxy using the telescope modules TM1, TM2, TM3, TM4, and TM6.

\subsection{Images}\label{sect_images}

\begin{figure*}[t]
\centering
\begin{minipage}{0.39\linewidth}
    \centering
    \includegraphics[width=\linewidth]{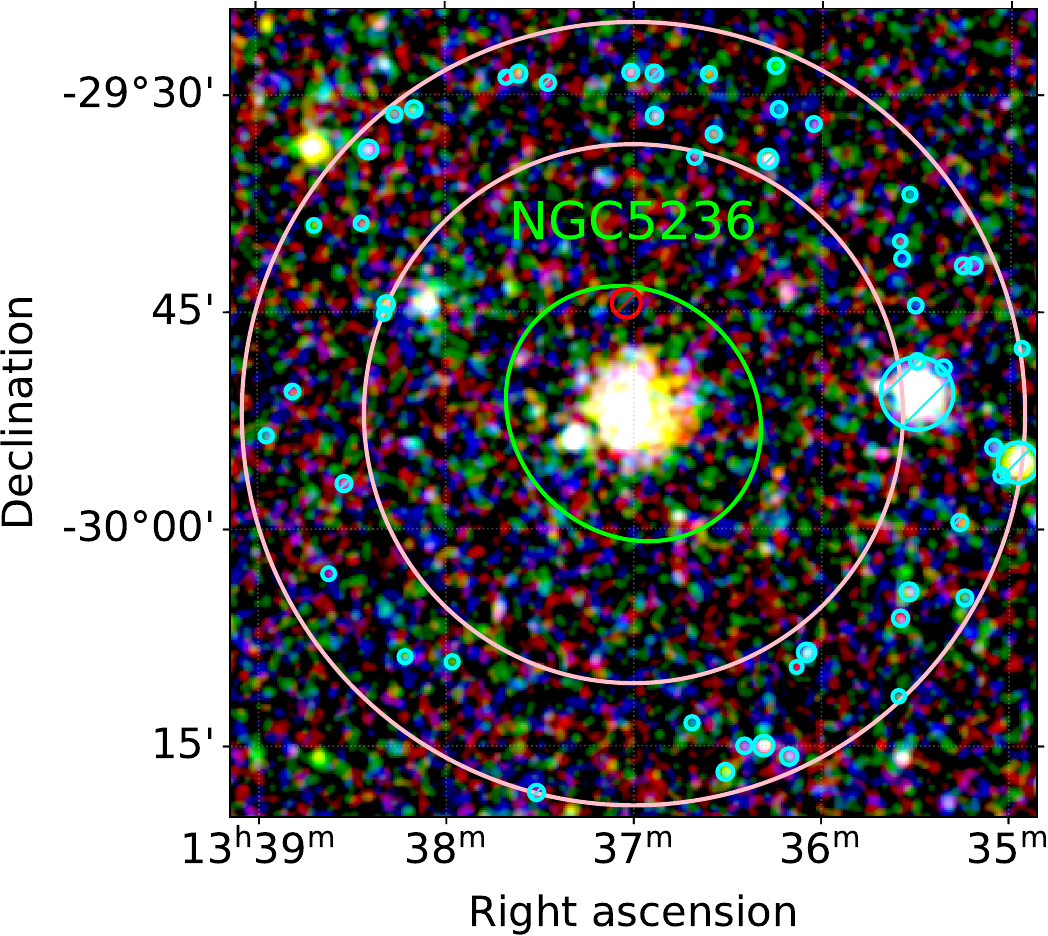}
\end{minipage}
\begin{minipage}{0.36\linewidth}
    \centering
    \includegraphics[width=\linewidth]{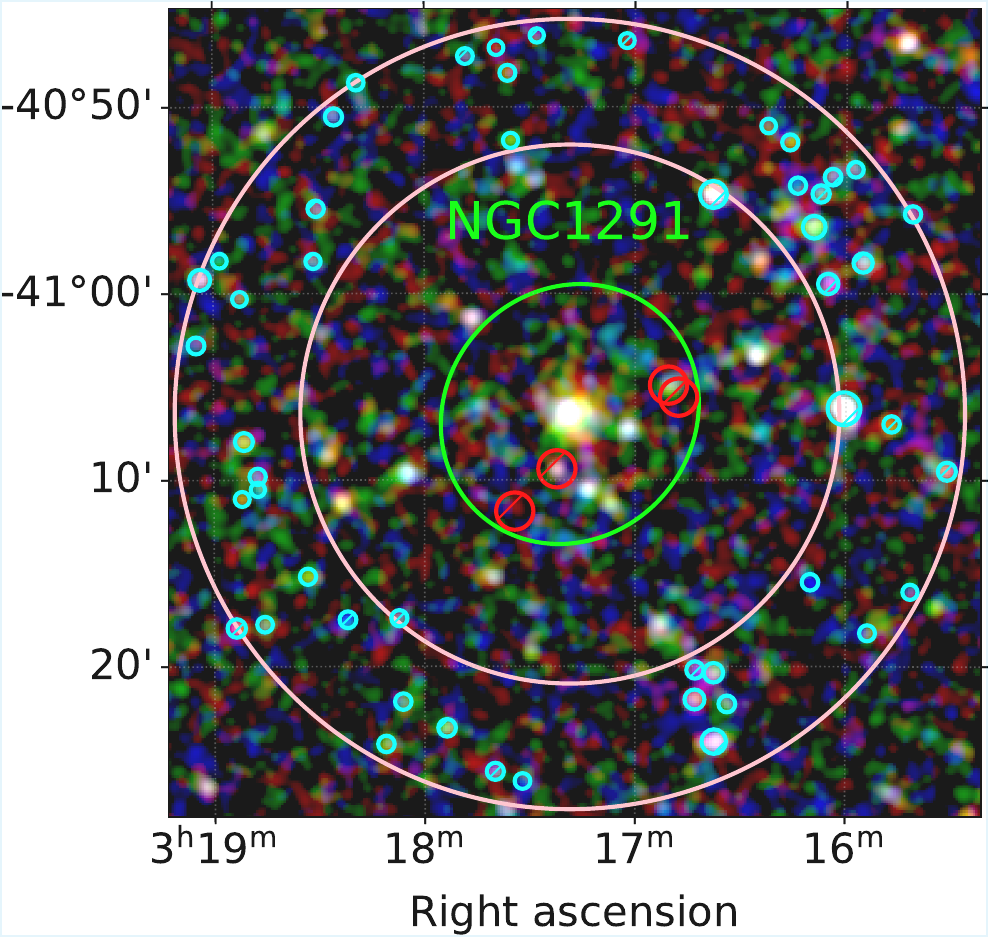}
\end{minipage}
\vspace{4pt}
\begin{minipage}{0.39\linewidth}
    \centering
    \includegraphics[width=\linewidth]{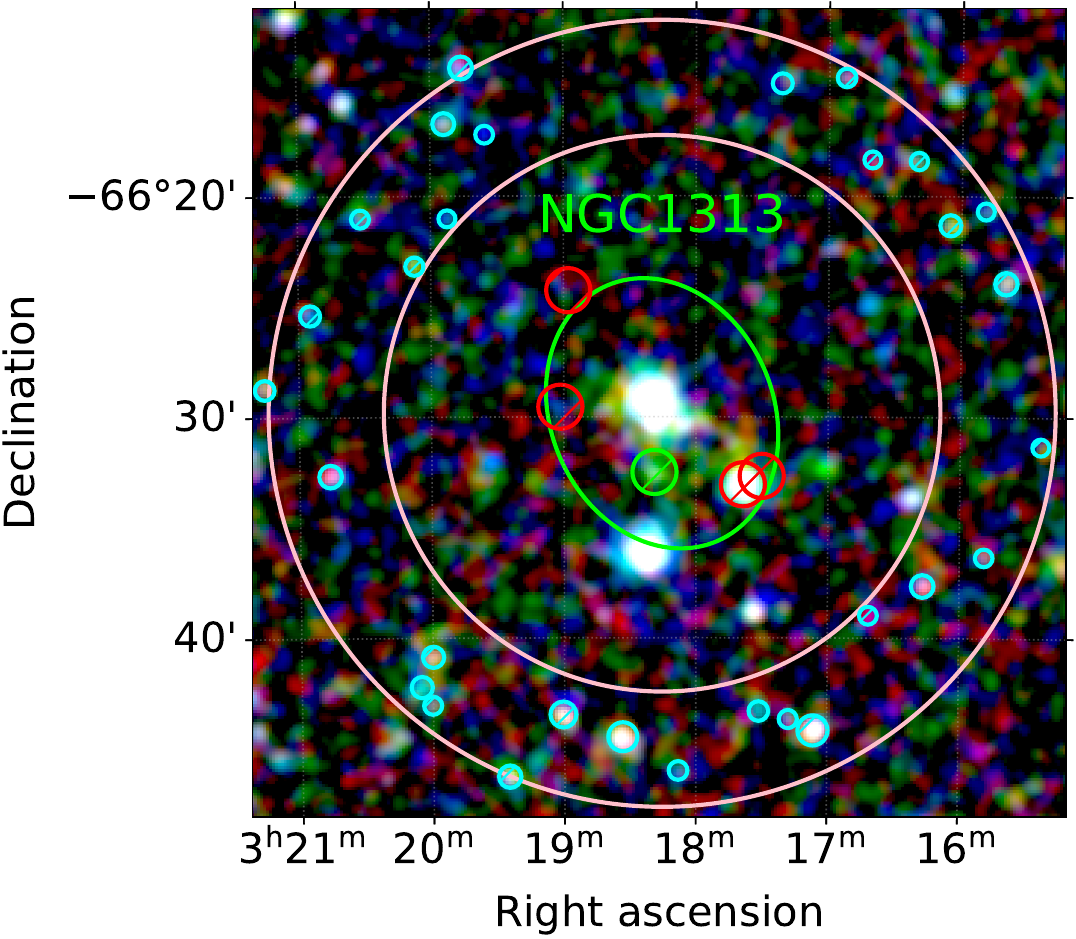}
\end{minipage}
\begin{minipage}{0.36\linewidth}
    \includegraphics[width=\linewidth]{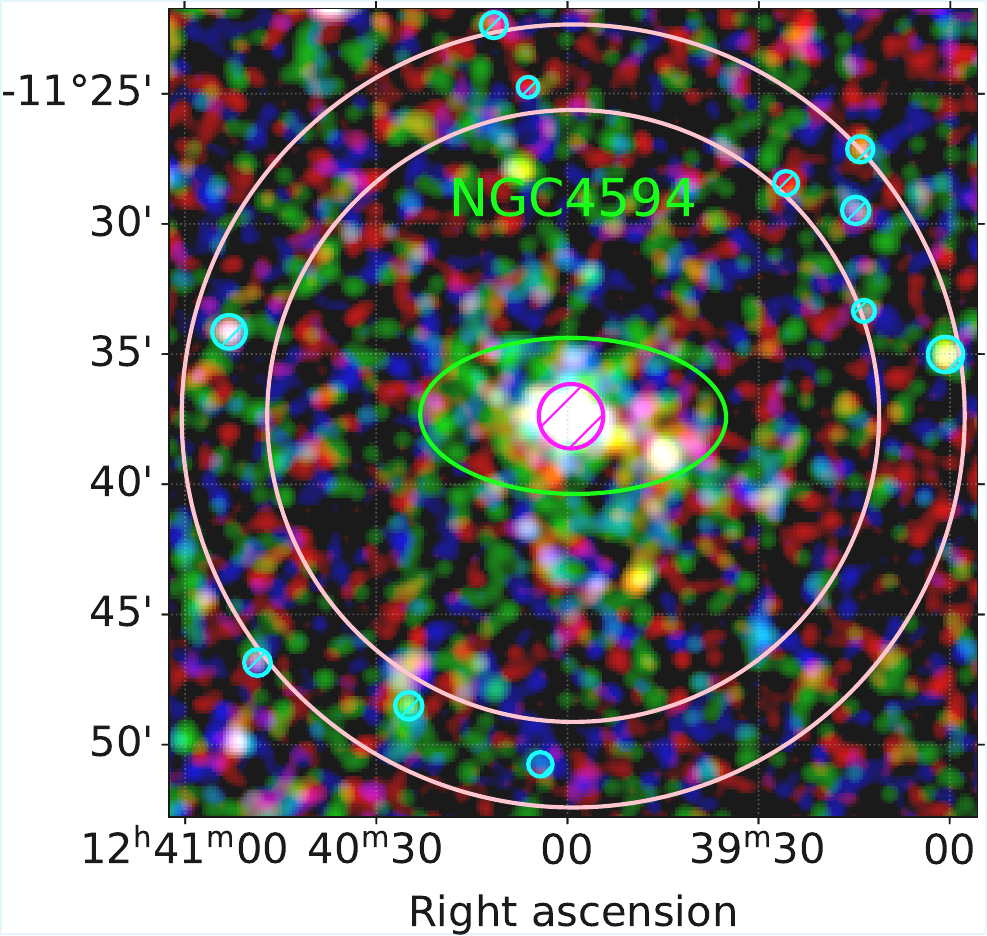}
\end{minipage}
\caption{eRASS:4 X-ray images of NGC~5236, NGC~1291, NGC~1313, and NGC~4594 in
$0.2$–$0.7$\,keV (red), $0.7$–$1.2$\,keV (green), and $1.2$–$5.0$\,keV (blue).
The green ellipses mark the optical extent of the galaxies (26.5\,mag Holmberg isophotes; \citealt{karachentsev}) and the pink
annuli the background extraction regions. The cyan circles mark excluded eRASS sources, the green circles mark foreground stars, and red circles mark background AGNs. The magenta circle in NGC 4594 marks the excluded nuclear region.}
\label{xray_imgs}
\end{figure*}

The images of NGC~5236, NGC~1291, NGC~1313, and NGC~4594 are shown in Fig.~\ref{xray_imgs}. The data are merged from the four all-sky survey scans (eRASS:4) and re-binned to a value of 160 sky pixels per image pixel, corresponding to an image scale of $8\arcsec \times 8\arcsec$ per pixel. Red, green, and blue colors correspond to the 0.2--0.7\,keV (soft), 0.7--1.2\,keV (medium), and 1.2--5.0\,keV (hard) bands, respectively. The images were centered on the galaxy positions from the Updated Nearby Galaxy Catalog \citep{karachentsev}. For each target, an elliptical source region was defined according to the optical extent of the galaxy (26.5\,mag isophote, green ellipses) using the semi-major axis $a$ and semi-minor axis $b$ from \citet{karachentsev}. The regions used to estimate the local background were defined as an annulus surrounding the source ellipse (pink annuli) with an inner radius at $2a$ and a background area five times larger than the source area. In these background regions, we marked all point sources from the eRASS:4 all-sky source catalog (cyan circles; eROSITA Consortium, in prep.). In the source regions, we marked background AGNs (red circles; WISE AGN candidates; \citealt{allwise}) and foreground stars (green circles; likely and possible identifications with HamStar; \citealt{hamstar}). The images reveal the distribution of the diffuse X-ray emission and X-ray point sources in the field of view. A detailed discussion of the individual galaxies is presented in Sect.~\ref{chap4}.

\subsection{Spectra}\label{sect_spectra}

We extracted merged eRASS:4 spectra for the galaxies using a \texttt{tophat} source extent model with a radius of $\sqrt{a \cdot b}$. For each galaxy, an additional local background spectrum was extracted using the annulus region. The spectra, as well as the associated Redistribution Matrix Files (RMFs) and Auxiliary Response Files (ARFs), were generated using the \texttt{srctool} task, which is part of \texttt{eSASS} \citep{brunner}.

Since we are interested in the contributions of diffuse hot gas and XRBs to the integrated X-ray emission, we did not mask point sources within the source regions. This also ensures a consistent treatment across the sample, since the contribution of individual ultraluminous X-ray sources (ULXs) cannot be reliably separated in more distant galaxies. We did, however, exclude sources that do not belong to the galaxies but would contaminate the spectra, i.e., background AGNs and foreground stars. These sources were identified via positional cross-matching with \texttt{TOPCAT}\footnote{\url{https://www.star.bris.ac.uk/~mbt/topcat/}}. In total, we found 65 AGN matches and 20 foreground star matches across the sample. For these sources, we applied circular exclusion regions with a radius of $1\arcmin$. In the background regions, we masked all point sources listed in the eRASS:4 catalog. This ensures that the estimated background is not biased by unrelated X-ray sources near the target, which could lead to overestimating of the local background. The masking radii for the excluded eRASS:4 sources were scaled empirically with count rate, following the power-law relation from \citet{haemmerich2026}.

A significant fraction of the galaxies in our sample are too faint for an individual spectral analysis. To investigate the diffuse emission of these systems in a statistical sense, we employ a stacking analysis. By combining the spectral data of multiple galaxies, stacking increases the statistical significance of the diffuse emission by reducing Poisson fluctuations, enabling spectral modeling in galaxy populations that would otherwise fall below the sensitivity limits of individual observations.

\subsection{Background model}

The eROSITA particle background is produced by cosmic rays interacting with the detector, and fluorescence emission from instrumental lines in the camera housing, mainly at ${\sim}1.5\,\mathrm{keV}$ (Al K$\alpha$) and $>6\,\mathrm{keV}$ (Fe K complex). We modeled the particle background with a combined filter-wheel-closed (FWC) model for TMs~1, 2, 3, 4, and~6, which was developed by the eROSITA consortium. An example background model decomposition for NGC 5236 can be viewed in the Appendix in Fig.~\ref{bkg_spectrum}.

For the X-ray background, we adopt an analytical model based on the results of \citet{yeung2024} and \citet{ponti}, who characterized the spectral properties of the soft Galactic X-ray background in the eROSITA all-sky and eFEDS surveys, respectively. The background is modeled as \texttt{constant(gauss + gauss + apec + tbabs(apec + apec + powerlaw))}. \texttt{apec} is a commonly used model for thermal plasma emission \citep{apec}. It quantifies the emission of a plasma in collisional ionization equilibrium accounting for, e.g., bremsstrahlung and recombination processes, as well as line emission from various elements.

The two Gaussian lines at 0.569\,keV and\,0.653 keV account for solar wind charge exchange (SWCX) emission from \ion{O}{vii} and \ion{O}{viii}. These lines originate from charge exchange processes between solar wind particles and the ISM.

The first, unabsorbed \texttt{apec} component represents emission from the Local Hot Bubble (LHB), a region of hot plasma ($kT{\sim}0.1\,\mathrm{keV}$) that surrounds the Sun. The temperature and abundance are fixed to $kT=0.1\,\mathrm{keV}$ and $Z=1\,Z_\odot$, which is consistent with values determined by \citet{yeung2024}.

The first absorbed \texttt{apec} models the emission from the circumgalactic medium (CGM). The CGM describes hot gas surrounding the disk of our Galaxy. Its temperature is allowed to vary between 0.15--0.25\,keV, with an abundance fixed at $0.1\,Z_\odot$. These values are consistent with other studies that find subsolar metallicities and plasma temperatures in the CGM of roughly $kT {\sim} 0.1\mbox{--}0.3\,\mathrm{keV}$ (e.g., \citealt{ponti,miller2015,henley2013}).

The second absorbed \texttt{apec} accounts for a hotter Galactic component, i.e., unresolved emission from stars, emission from superbubbles, or a higher-temperature halo component. It is modeled with a fixed temperature $kT=0.7$ keV and abundance $Z=1\,Z_\odot$, in accordance with \citet{yeung2024}.

The absorbed \texttt{powerlaw} represents the cosmic X-ray background (CXB), i.e., the combined emission of distant AGNs, galaxies, and galaxy clusters. It is modeled by a power-law with fixed photon index $\Gamma = 1.46$, a standard value for the CXB (e.g. \citealt{cappelluti2017}).

The Galactic absorption is described by the Tuebingen-Boulder ISM absorption model (\texttt{TBabs} model in \texttt{XSPEC}; \citealt{wilms}). The values for the Galactic \ion{H}{i} column density are obtained from the HI4PI HI survey (\citealt{HI4PI}). The model is multiplied by a constant factor to allow for an adjustment of the extraction area.

\subsection{Source model}

According to \textit{Chandra}, \textit{XMM-Newton} or \textit{Suzaku} studies (e.g., \citealt{intr7,intr8}) of diffuse X-ray emission in nearby galaxies (e.g., the Magellanic Clouds or M31) and our own Galaxy, the diffuse emission in most galaxies is consistent with thermal plasma emission at temperatures of $kT \sim \SI{0.2}{keV}$. This emission component most likely originates from the hot ISM in equilibrium and unresolved stellar sources. In addition, many galaxies also display hotter areas such as \ion{H}{ii} regions, supernova remnants (SNRs), superbubbles, and emission from flare stars or active binaries. The contribution of these regions can be modeled by a second thermal plasma component with a temperature of $kT > \SI{0.5}{keV}$. In addition, we expect a significant contribution to the spectrum by (unresolved) XRBs. \citet{mineo2} studied a sample of 29 nearby star-forming galaxies using \textit{Chandra} data, and found that, on average, ${\sim}30\mbox{--}40\,\%$ of the galaxies' 0.5--2\,keV luminosity from diffuse emission is produced by faint compact sources. They derive a best-fitting HMXB model with two thermal components and a power-law with a slope of $\Gamma = 1.8$ and intrinsic absorption with $N_\mathrm{H} = \SI{4.9e21}{cm^{-2}}$.

Due to the limited photon statistics of the eROSITA data, we apply a single \texttt{apec} model to characterize emission from the hot gas, and one power-law component to quantify the XRB contribution. The plasma temperature should therefore be regarded as a mean effective temperature of the dominant thermal component, and the power-law normalization as a measure of the total XRB-related emission. We add a second \texttt{TBabs} component to account for intrinsic absorption within the galaxies. The total spectral model is therefore \texttt{constant(background + tbabs(tbabs(apec + powerlaw)))}.

\subsection{Spectral analysis}

The spectral analysis was performed using the X-Ray Spectral Fitting Package \texttt{XSPEC}\footnote{\url{https://heasarc.gsfc.nasa.gov/docs/software/xspec/}} (version~12) in the 0.3--9\,keV energy range. The data were not binned and fitted using the Cash statistic (\texttt{cstat}, \citealt{cash1979}). The background spectra were fitted using the ARFs generated for the background regions, after which the shape of the background spectrum was frozen, while the normalizations were left free. For each spectrum, we determined a scaling factor $\gamma_\mathrm{i}$ to rescale the background emission to the source region, 

\begin{equation}\label{gamma}
    \gamma_\mathrm{i} = \frac{B_\mathrm{src,i}E_\mathrm{src,i}}{B_\mathrm{bkg,i}E_\mathrm{bkg,i}}, \hspace{20pt} \gamma_\mathrm{stack} =\frac{\sum_i B_{\mathrm{src,i}}\,E_{\mathrm{src,i}}}
     {\sum_i B_{\mathrm{bkg,i}}\,E_{\mathrm{bkg,i}}},
\end{equation}

where $B$ denotes the \texttt{BACKSCAL} value stored in the spectral header and $E$ the exposure time of the source and background regions. In our spectra, \texttt{BACKSCAL} represents the effective extraction area of the corresponding region. This factor was applied to the \texttt{constant} component of the source model, after which the source and background spectra were modeled simultaneously. The element abundances of the source model were fixed to solar values for all galaxies. The parameter uncertainties correspond to the 90\,\% confidence intervals of one specific parameter. They were determined using the \texttt{error} command in \texttt{XSPEC}.

For the stacking analysis, stacked source and background spectra were constructed for each subsample by summing the counts per energy channel and the corresponding exposure times of each individual galaxy $i$. The ARFs for both the source and background spectra were averaged using weights given by $B_\mathrm{i}\cdot E_\mathrm{i}$. The scaling factor $\gamma_\mathrm{stack}$ for the stacked spectra is given by (Eq.~\ref{gamma}).

For the Galactic $N_\mathrm{H}$, we used an exposure-weighted mean of the individual Galactic column densities derived from the HI4PI survey \citep{HI4PI}. During the fit, the normalizations of the two SWCX Gaussian components consistently approached zero, indicating that any SWCX emission is diluted when averaging over different SWCX conditions. The SWCX normalizations were therefore fixed to zero.

In cases where individual model parameters were not well constrained, we fixed them to baseline values motivated either by literature values or, where possible, by the behavior of the fit and the confidence contours. To assess the resulting systematic uncertainties, we repeated the fits for a set of fixed parameter values. For each fit, we computed the $90\,\%$ parameter confidence intervals and quoted the full parameter range by taking the lowest and highest bounds from all explored models.

\subsection{Fluxes and luminosities}\label{sect_fluxes}

We determined the absorption-corrected thermal (\texttt{apec}) and total (\texttt{apec+powerlaw}) fluxes of the individually modeled galaxies and the stacked spectra using \texttt{cflux}, an \texttt{XSPEC} convolution model that integrates the model flux over a specified energy band and treats the flux itself as a fit parameter. Fluxes were measured in the 0.5--2\,keV and 0.5--8\,keV bands. 

For each individual galaxy, the flux was estimated from the count-rate-to-flux conversion of the corresponding stacked spectrum

\begin{equation}
    F_\mathrm{i} = \frac{R_\mathrm{i}}{R_\mathrm{s}}\,F_\mathrm{s}, \hspace{20pt} R_\mathrm{i} = \frac{C_\mathrm{i,src} - \gamma_\mathrm{i}\,C_\mathrm{i,bkg}}{E_\mathrm{i,src}},
\end{equation}

where $R_\mathrm{i}$ is the net count rate of galaxy $i$, obtained by subtracting the scaled background counts $C_\mathrm{i,bkg}$ from the source counts $C_\mathrm{i,src}$, and $R_\mathrm{s}$ and $F_\mathrm{s}$ are the net count rate and absorption-corrected flux of the corresponding stacked spectrum. This method was applied separately to the total and thermal fluxes. Luminosities were then calculated via $L_\mathrm{i} = 4\pi d_\mathrm{i}^2F_\mathrm{i}$ using the adopted galaxy distances from \citet{karachentsev}. This procedure assumes that the stacked spectral model provides an average count-rate-to-flux conversion, representative of the galaxies in the subsample. Individual galaxies may deviate from this average.

The uncertainties were propagated using a Monte Carlo approach with $2 \cdot 10^4$ realizations. Source and background counts were resampled assuming Poisson statistics with \texttt{numpy.random.default\_rng().poisson}. For the stacked spectra, the sampled net counts were obtained by summing the Monte Carlo realizations of the individual galaxies; the count rates were derived from the summed exposure.

The uncertainties of the stacked fluxes were taken from the $90\,\%$ confidence range of the \texttt{cflux} parameter in \texttt{XSPEC}. For the Monte Carlo propagation, we resampled these fluxes via an asymmetric normal distribution using \texttt{numpy.random.default\_rng().normal}. The widths of the sampling distributions were then matched to the confidence bounds using the inverse cumulative normal distribution \texttt{statistics.NormalDist().inv\_cdf}. For detected components, we matched the distribution to the lower and upper $90\,\%$ confidence bounds; for components with only upper limits we used a one-sided distribution consistent with the $90\,\%$ upper bound. These sampled stack fluxes and count rates were then used to derive sampled count-rate-to-flux conversion factors, which we applied to the sampled count rates of the individual galaxies.

The resulting flux and luminosity distributions were characterized by their median and percentile-based confidence bounds, computed with \texttt{numpy.percentile}. When the final, lower $90\,\%$ error bound of a given galaxy flux remained positive, this flux was treated as a detection. Otherwise, we report a $90\,\%$ upper limit. 

We used the extended-source sensitivity of eFEDS \citep{liu2022} to get a rough estimate for the soft-band sensitivity limit $F_\mathrm{lim}$ of our data. \citet{liu2022} reported a soft-band flux limit of ${\sim}10^{-14}\,\mathrm{erg\,s^{-1}\,cm^{-2}}$ within a $1\arcmin$ aperture for eFEDS, which has an exposure comparable to the full-depth eROSITA all-sky survey. Our average extraction region (an ellipse with $a{\sim}2\arcmin$, $b{\sim}1\arcmin$) has an area of $A {\sim} 2\pi\,\mathrm{arcmin^{2}}$; twice the area of a $1\arcmin$-radius aperture. Assuming Poisson statistics, $F_\mathrm{lim} \propto \sqrt{A/t}$, with an exposure time of eRASS:4 that is approximately two times lower compared to the reference depth of eFEDS, we estimate a sensitivity of $F_\mathrm{lim}{\sim}2\cdot10^{-14}\,\mathrm{erg\,s^{-1}\,cm^{-2}}$.

\subsection{$L_\mathrm{X}-\mathrm{SFR}$ fitting}

We investigate the scaling relation between the X-ray luminosity and the SFR of the star-forming galaxies in our sample. Before fitting the relation, we inspected galaxies with $\mathrm{SFR}<10^{-3}\,\mathrm{M_{\odot}\,yr^{-1}}$ to assess whether these very low catalog SFRs are plausible. We find that four objects, VV747, KKH58, CGCG-095-078, and IC0559, show optical properties consistent with blue, low-metallicity starburst systems, for which standard mid-IR/optical SFR calibrations can underestimate the true SFR. We additionally find IC 2038, a star-forming spiral galaxy, as having an unusually low HECATEv2 SFR of $8.4\cdot10^{-6}\,\mathrm{M_{\odot}\,yr^{-1}}$. A comparison with the literature revealed that this value disagrees by more than three orders of magnitude with the SFR of $(2.5\pm0.2)\cdot10^{-2}\,\mathrm{M_{\odot}\,yr^{-1}}$ derived by \citet{rampazzo2021}. We therefore exclude these five galaxies from the $L_\mathrm{X}-\mathrm{SFR}$ fit. For NGC 5236, we use the luminosity derived from the individual fit, since its more than 1000 net counts were dominating the medium-sSFR stacked spectrum; it was therefore excluded from the stack, but kept in the $L_\mathrm{X}-\mathrm{SFR}$ fit. The final fit uses 122 objects: 22 high-, 47 medium-, and 53 low-sSFR galaxies.

The total luminosity (\texttt{apec} + \texttt{powerlaw} in 0.5--8\,keV band) and the luminosity of the thermal component (only \texttt{apec}, in 0.5--2\,keV band) were fitted as a function of the SFR using the model
\begin{equation}\label{eq_log_fit}
    \log L_\mathrm{X} = A + \alpha\, \log \mathrm{SFR}.
\end{equation}
We included an intrinsic scatter term $\sigma_\mathrm{int}$ in log-space to account for astrophysical variation and fitted detections and upper limits with a maximum-likelihood approach. Asymmetric luminosity uncertainties were symmetrized in log-space. The details of the likelihood function are given in Appendix~\ref{appendix_like}. The fit parameters $A$, $\alpha$, and $\sigma_\mathrm{int}$ were determined by minimizing the negative log-likelihood using \texttt{scipy.optimize.minimize} in \texttt{Python} (version 3.11.14).

\section{Results}\label{chap4}

\begin{table*}
\caption{Fit results and absorption-corrected fluxes for the individually fitted galaxies.}   
\label{tab_res1}
\centering
\begin{tabular}{llcccc}
\hline\hline
        Parameter                 & Unit                                      & NGC 5236               & NGC 1291                & NGC 1313                & NGC 4594                 \\\hline
        Cts                       &                                           & 1264                   & 412                     & 1353                     & 183                      \\
        Distance                  & Mpc                                       & 4.9                    & 8.8                     & 4.1                     & 9.3                      \\
        Hubble-Type               &                                           & 5.0                    & 0.1                     & 7.0                     & 1.1                      \\\vspace{5pt}
        SFR                       & $\mathrm{M_\odot\,yr^{-1}}$               & 4.0                    & 0.16                    & 0.71                    & 0.33                     \\
        $M_{\ast}$                & $10^{10}\,\mathrm{M_\odot}$               & $2.4$                  & $5.0$                   & $0.19$       & $12.4$        \\
        cstat/dof                 &                                           & 1.11                   & 1.04                    & 1.09                    & 0.70                     \\
        $N_\mathrm{H,Gal}$        & $10^{22}\,\mathrm{cm^{-2}}$               & 0.04                   & 0.014                   & 0.07                    & 0.04                     \\
        $N_\mathrm{H,int}$        & $10^{22}\,\mathrm{cm^{-2}}$               & $< 0.08$               & $0.14^{+0.23}_{-0.13}$  & $0$                     & $< 0.18$                      \\\vspace{2pt}
        $kT$                      & keV                                       & $0.29^{+0.05}_{-0.04}$ & $0.40^{+0.34}_{-0.18}$  & $0.88^{+0.11}_{-0.10}$  & $> 0.60$                 \\\vspace{2pt}
        $\text{norm}_\text{apec}$ & $10^{-4}\,\mathrm{cm^{-5}}$               & $44^{+15}_{-12}$       & $4.9^{+7.6}_{-3.6}$     & $10.3^{+3.3}_{-3.0}$     & $4.1^{+3.2}_{-2.8}$ \\\vspace{2pt}
        $\Gamma$                  &                                           & $2.0^{+0.6}_{-0.4}$    & $3.0^{+1.6}_{-1.1}$     & $1.1^{+0.2}_{-0.2}$     & $2.2^{+1.3}_{-0.5}$      \\\vspace{2pt}
        $\text{norm}_\text{PL}$   & $10^{-4}\,\mathrm{cts\,keV^{-1}\,cm^{-2}\,s^{-1}}$ 
                                                                              & $36^{+10}_{-6}$        & $11^{+9}_{-5}$          & $24.6^{+2.9}_{-2.9}$     & $7.7^{+6.3}_{-2.7}$ \\
        $F_\text{apec}^{\mathrm{0.5\mbox{--}2\,keV}}$  & $10^{-13}\,\mathrm{erg\,s^{-1}\,cm^{-2}}$
                                                                              & $10.3^{+3.2}_{-1.9}$   & $1.4^{+1.5}_{-0.7}$     & $3.2^{+0.6}_{-0.6}$     & $1.3^{+0.9}_{-0.6}$ \vspace{2pt}\\
        $F_\text{tot}^{\mathrm{0.5\mbox{--}2\,keV}}$   & $10^{-13}\,\mathrm{erg\,s^{-1}\,cm^{-2}}$     
                                                                              & $26.9^{+5.3}_{-2.4}$   & $6.5^{+6.5}_{-2.1}$     & $13.8^{+0.8}_{-0.8}$     & $4.5^{+0.8}_{-0.8}$    \vspace{2pt}\\
        $F_\text{tot}^{\mathrm{0.5\mbox{--}8\,keV}}$   & $10^{-13}\,\mathrm{erg\,s^{-1}\,cm^{-2}}$     
                                                                              & $43.1^{+8.2}_{-5.7}$   & $7.8^{+5.3}_{-1.8}$     & $50.2^{+8.5}_{-7.1}$     & $6.9^{+2.8}_{-1.8}$    \vspace{2pt}\\        
\hline
\end{tabular}
\begin{tablenotes}
\footnotesize
\item \textit{Notes.} Counts in 0.3--9\,keV, Distance from \citet{karachentsev}, Hubble-Type, SFRs and stellar masses from HECATEv2 \citep{hecatev2}, Galactic and intrinsic hydrogen column densities $N_\mathrm{H,Gal}$ and $N_\mathrm{H,int}$, plasma temperature $kT$, power-law index $\Gamma$, normalizations of the \texttt{apec} and \texttt{powerlaw} models, Fluxes $F$ for different bands and model components.
\end{tablenotes}
\end{table*}

First, we analyze the two brightest star-forming galaxies, NGC~5236 and NGC~1313, the brightest early-type galaxy, NGC~1291, and an AGN control galaxy NGC~4594 (which is not part of our galaxy sample), individually. These galaxies span a range of evolutionary stages and conditions. They provide reference cases for interpreting the stacked spectra discussed below. The spectra of the four galaxies are shown in Fig.~\ref{spectra} in the Appendix.

\subsection{NGC 5236}

NGC 5236 (M83) is a face-on ($i =26^{\circ}$; \citealt{n5236_9}) barred spiral galaxy at a distance of ${\sim}\SI{4.9}{Mpc}$ \citep{karachentsev}. It is the brightest galaxy and has the highest SFR in our sample (${\sim} 4.0\,\mathrm{M_{\astrosun}\,yr^{-1}}$). The X-ray image (Fig.~\ref{xray_imgs}, top left) reveals bright, extended emission and numerous point sources distributed across the galaxy. The eROSITA spectrum (Fig.~\ref{spectra}, top left) is well described by a thermal plasma with a plasma temperature of $kT = 0.29^{+0.05}_{-0.04}\,\mathrm{keV}$ and a power-law component with a photon index of $\Gamma = 2.0^{+0.6}_{-0.4}$. The derived plasma temperature is consistent with the value of $kT = 0.24^{+0.02}_{-0.02}\,\mathrm{keV}$, derived from \textit{XMM-Newton} observations by \citet{owen2009}. NGC 5236 serves as a representative example of a star-forming galaxy in which the spectrum is dominated by soft thermal emission, with an additional harder component from unresolved compact sources.

\subsection{NGC 1291}

NGC 1291 is a nearly face-on, early-type spiral galaxy at a distance of ${\sim}8.8\,\mathrm{Mpc}$ \citep{karachentsev}, viewed at an inclination of $i {\sim} 35\,^{\circ}$ \citep{ngc1291_53}. It is one of the early-type systems in our sample and has a comparatively low SFR of ${\sim}0.16\,\mathrm{M_{\odot}\,yr^{-1}}$. The X-ray image (Fig.~\ref{xray_imgs}, top right) shows bright emission concentrated in the center, with only weak large-scale diffuse emission distributed across the disk. The spectrum (Fig.~\ref{spectra}, top right) is dominated by a power-law continuum over most of the fitted energy range, while the soft thermal component is relatively weak. It is well described by a thermal plasma with $kT = 0.40^{+0.34}_{-0.18}\,\mathrm{keV}$ and a power-law component with $\Gamma = 3.0^{+1.6}_{-1.1}$. The plasma temperature is consistent with the result of \citet{irwin2002}, who find $kT{\sim}0.33\mbox{--}0.34\,\mathrm{keV}$ using a \textit{Chandra} observation. NGC 1291 serves as a representative example of an early-type system in which the diffuse emission is weaker and the integrated spectrum is dominated by a hard, nonthermal component.

\subsection{NGC 1313}

NGC 1313 is an intermediately inclined ($i = 41^{\circ}$) late-type barred spiral galaxy at a distance of ${\sim}4.1\,\mathrm{Mpc}$ \citep{karachentsev}. It has a moderate star-forming activity with a SFR of ${\sim}0.71\,\mathrm{M_{\odot}\,yr^{-1}}$. The galaxy hosts two ULXs, NGC 1313 X-1 and X-2. The X-ray image (Fig.~\ref{xray_imgs}, bottom left) shows extended soft X-ray emission across the central region, with noticeable structure in the south-western part of the galaxy. The spectrum (Fig.~\ref{spectra}, bottom left) is described by a thermal plasma with $kT = 0.88^{+0.11}_{-0.10}\,\mathrm{keV}$ and a power-law component with $\Gamma = 1.1^{+0.2}_{-0.2}$. The total absorption-corrected flux in the 0.5--8\,keV band is $5.0^{+0.9}_{-0.8}\cdot10^{-12}\,\mathrm{erg\,s^{-1}\,cm^{-2}}$. NGC 1313 illustrates that even in a galaxy with very luminous point sources, the diffuse component can still be spectrally constrained.

\subsection{NGC 4594}

NGC 4594 is an early-type lenticular galaxy viewed nearly edge-on ($i \sim 79^{\circ}$; \citealt{ngc4594_4}) at a distance of ${\sim}9.3\,\mathrm{Mpc}$ \citep{karachentsev}. It is one of the most massive galaxies in the local volume and hosts a low-luminosity AGN with radio jets \citep{ngc4594_21,ngc4594_61}. NGC 4594 serves as a control case for residual nuclear contamination. The bright central region was excluded in the spectral analysis. The X-ray image (Fig.~\ref{xray_imgs}, bottom right) shows strongly concentrated X-ray emission in the central part of the galaxy, with weaker diffuse emission outside of the masked nucleus. The spectrum (Fig.~\ref{spectra}, bottom right) is harder than those of the star-forming galaxies and lacks a soft excess. It is described by a power-law component with $\Gamma = 2.2^{+1.3}_{-0.5}$, while the thermal component is only weakly constrained with a temperature of $kT > 0.60\,\mathrm{keV}$. NGC 4594 provides an example of a system in which the X-ray output is strongly affected by nuclear processes and the X-ray luminosity does not mainly arise from star-formation-related processes.

\subsection{Stacked spectra}\label{sect_stack}

\begin{figure*}
    \centering
    \begin{tabular}{cc}
        \includegraphics[width=.42\textwidth]{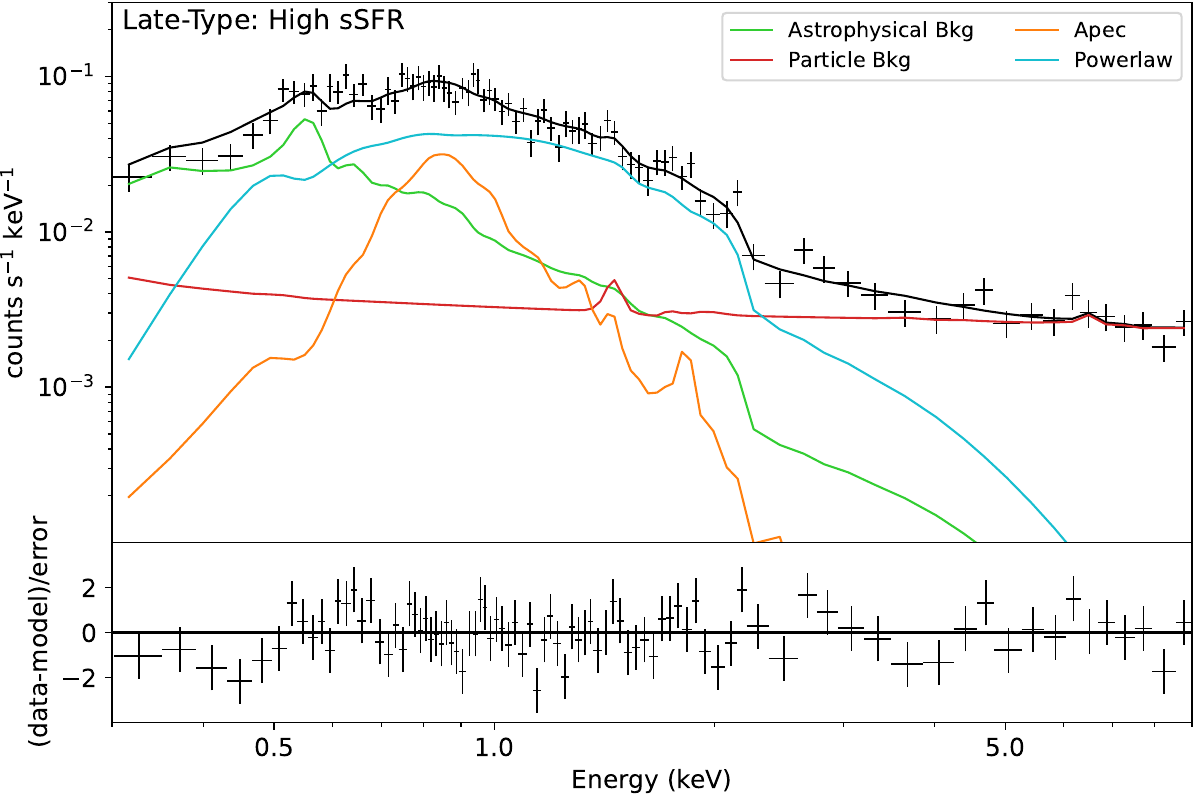}&
        \includegraphics[width=.42\textwidth]{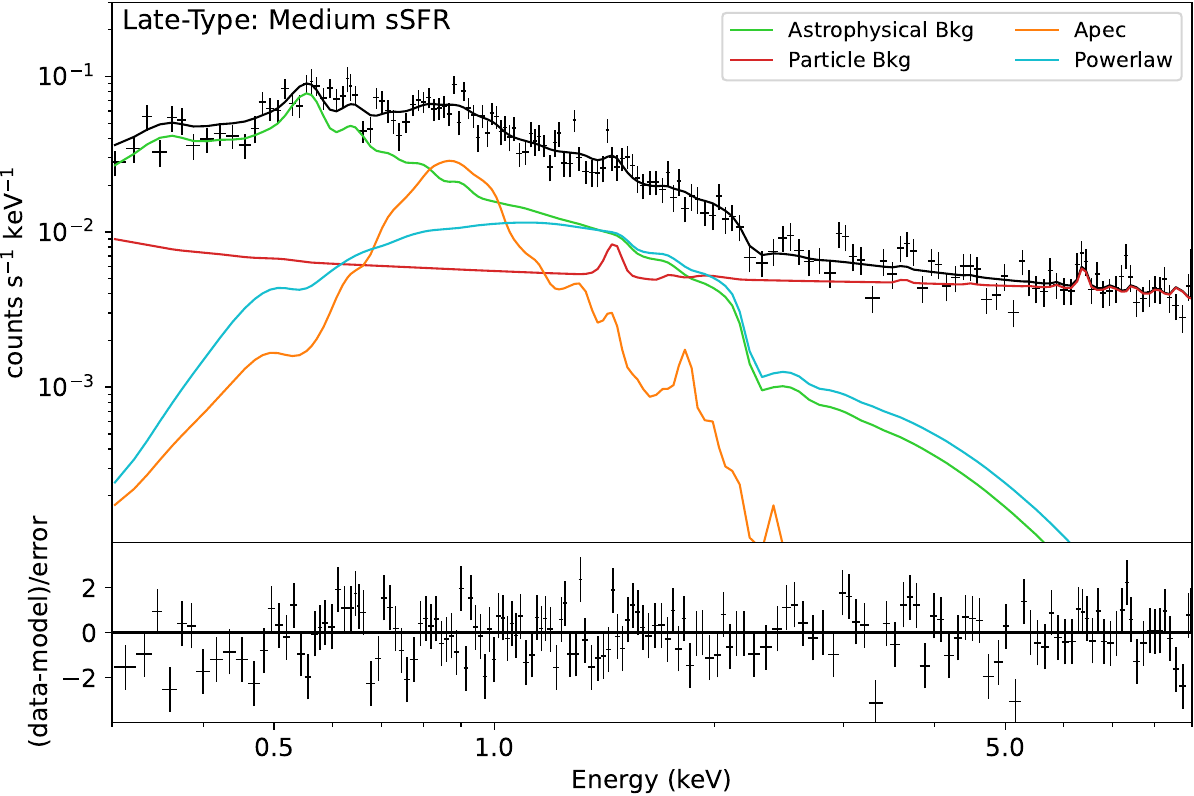}\\
        \includegraphics[width=.42\textwidth]{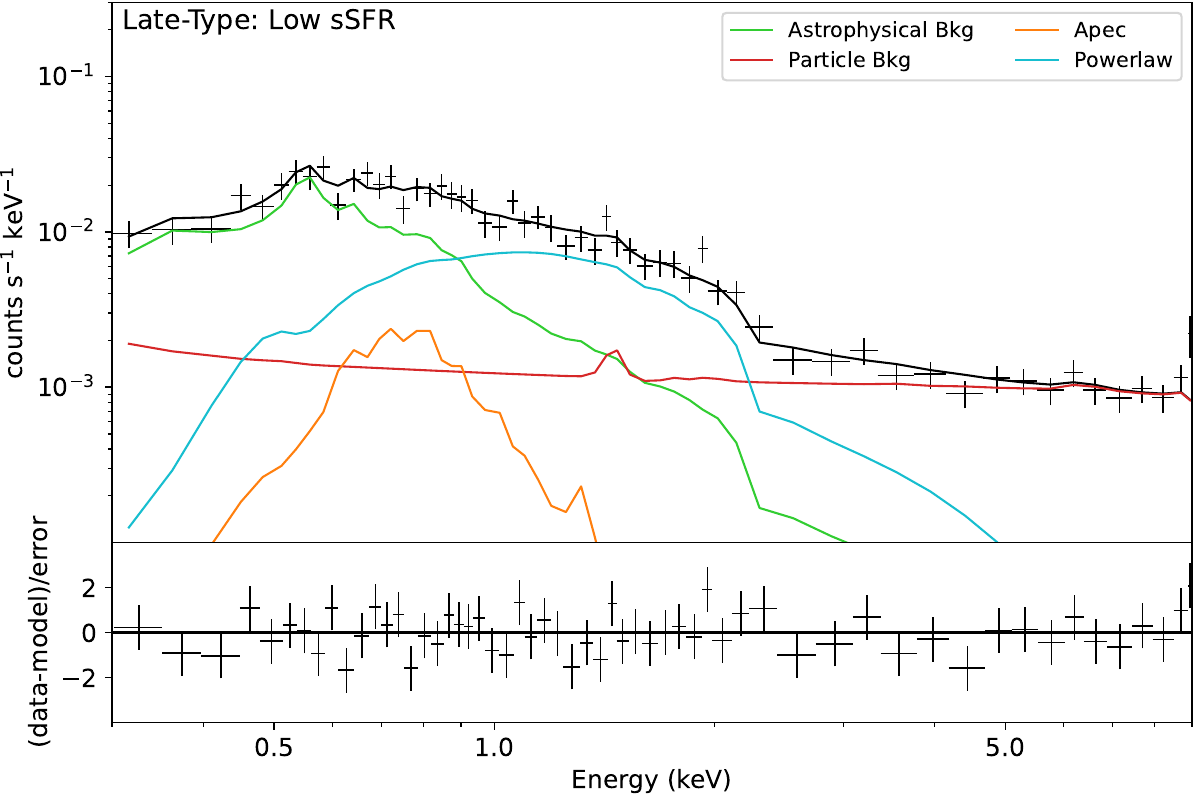}&
        \includegraphics[width=.42\textwidth]{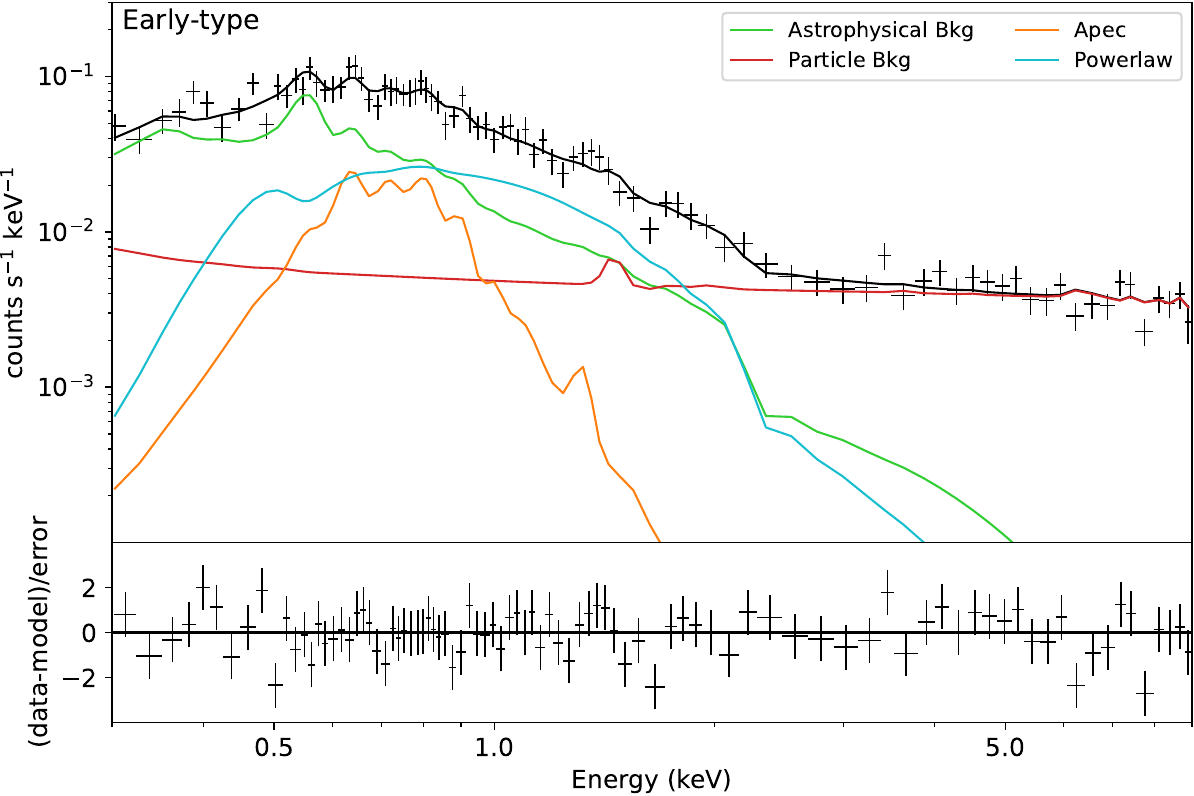}\\
    \end{tabular}
      \caption{Stacked eRASS:4 spectra of the high (upper left), medium (upper right), and low (lower left) sSFR galaxy subsamples. The spectrum of the early-type galaxies is shown in the lower right panel. The black lines show the full spectral model, the colored lines show the contributions of the different model components.}
         \label{spec_stacked}
\end{figure*}

The fit results of the stacked spectra are listed in Tab.~\ref{tab_res3}. The spectrum of the high-sSFR galaxies (Fig.~\ref{spec_stacked}, top left) shows a clear excess below ${\sim}1\,\mathrm{keV}$, together with a broad nonthermal continuum. The spectrum is well described by a thermal plasma with $kT=0.76^{+0.13}_{-0.17}\,\mathrm{keV}$ and a power-law component with a photon index of $\Gamma=2.1^{+0.6}_{-0.5}$.

The spectrum of the medium-sSFR galaxies (Fig.~\ref{spec_stacked}, top right) also shows a clear soft excess below ${\sim}1\,\mathrm{keV}$ and a continuum extending to higher energies. The thermal component is well fitted at $kT = 0.82^{+0.08}_{-0.09}\,\mathrm{keV}$, while the photon index is less tightly constrained with an upper limit of $\Gamma < 2.2$.

The spectrum of the low-sSFR galaxies (Fig.~\ref{spec_stacked}, bottom left) is harder than the high- and medium-sSFR spectra and is dominated by the power-law component with a photon index of $\Gamma = 1.8^{+0.9}_{-0.7}$. A weak soft excess is present below ${\sim}1\,\mathrm{keV}$, but the plasma temperature cannot be constrained by the data. To assess the systematic uncertainty introduced by the fixed model parameters, we repeated the fit on a grid of fixed values with $kT = 0.2, 0.3, 0.5\,\mathrm{keV}$, motivated by the mean, cool temperature reported by \citet{mineo2}, $\left\langle kT \right\rangle = 0.24\,\mathrm{keV}$, for the diffuse emission in nearby star-forming galaxies. The quoted statistical uncertainties correspond to the $90\,\%$ confidence intervals of the adopted reference fit with $kT = 0.3\,\mathrm{keV}$. The systematic error ranges were determined from the full scatter of parameter and flux error bounds across all three model configurations. For all assumed temperatures, the normalization of the thermal component is only constrained by an upper limit. The most conservative upper limit is obtained from the $kT = 0.2\,\mathrm{keV}$ model and is used to estimate the thermal flux $F_\mathrm{apec}^{0.5-2\,\mathrm{keV}} < 5.8\cdot10^{-14}\,\mathrm{erg\,s^{-1}\,cm^{-2}}$. The total flux in 0.5--8\,keV corresponds to $1.0^{+0.4}_{-0.2}\cdot10^{-13}\,\mathrm{erg\,s^{-1}\,cm^{-2}}$ including both statistical and systematic uncertainties.

The spectrum of the early-type galaxies (Fig.~\ref{spec_stacked}, bottom right) contains both a cool thermal component and a power-law continuum. The fit yields $kT = 0.26^{+0.35}_{-0.11}\,\mathrm{keV}$ and $\Gamma=3.5^{+1.8}_{-1.4}$.

\subsection{Fluxes and luminosities}

\begin{figure*}
    \centering
    \begin{tabular}{cc}
        \includegraphics[width=.94\textwidth]{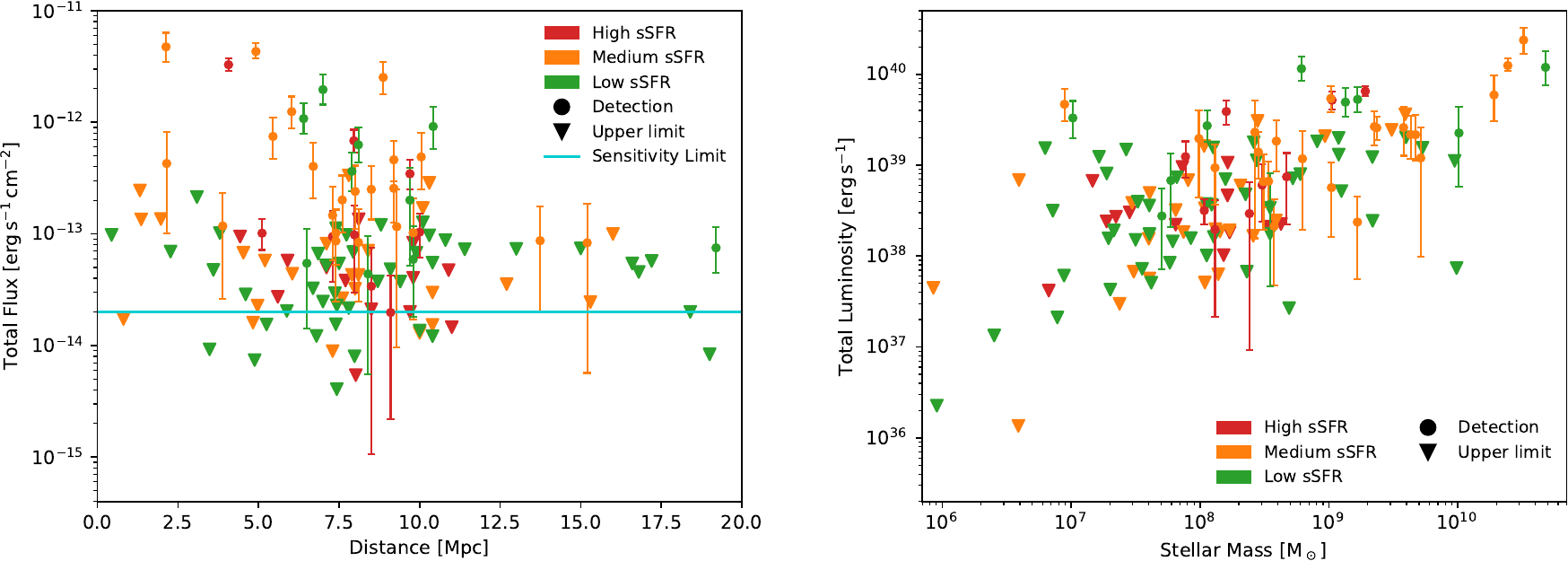}
    \end{tabular}
         \caption{Total 0.5--2\,keV flux over distance (left) and total 0.5--2\,keV luminosity over stellar mass (right) for the star-forming galaxies. High-, medium-, and low-sSFR galaxies are shown in red, orange, and green. Circles mark detections, triangles mark upper limits. The cyan line marks the approximate eRASS:4 sensitivity limit of ${\sim}2\cdot10^{-14}\,\mathrm{erg\,s^{-1}\,cm^{-2}}$. Distances were taken from \citet{karachentsev}, stellar masses from HECATEv2 \citep{hecatev2}.}
         \label{pl_flux}
\end{figure*}

We plot the total fluxes and luminosities in the soft band, corrected for Galactic absorption, in Fig.~\ref{pl_flux}. Sources for which the lower 90\,\% flux error is consistent with zero are shown as upper limits; all other sources are treated as detections. 

The left panel shows the total 0.5--2\,keV flux as a function of distance. The detected sources are mostly located above the adopted eRASS:4 sensitivity limit of ${\sim}2\cdot10^{-14}\,\mathrm{erg\,s^{-1}\,cm^{-2}}$, while the upper limits are concentrated near and below this threshold. The detected fluxes cover a range of approximately $10^{-14}$ to a few $10^{-12}\,\mathrm{erg\,s^{-1}\,cm^{-2}}$, with the brightest sources located at distances below ${\sim}10\,\mathrm{Mpc}$.

The right panel shows the total 0.5--2\,keV luminosity as a function of stellar mass. The luminosities span four orders of magnitude, from ${\sim}10^{36}$ to ${\sim}10^{40}\,\mathrm{erg\,s^{-1}}$. The brightest sources are found among the most massive galaxies, at $M_{\ast} \gtrsim 10^{9}\,\mathrm{M_{\odot}}$, and are mostly detections. At intermediate stellar masses, $M_{\ast} {\sim} 10^{7} \mbox{--} 10^{9}\,\mathrm{M_{\odot}}$, the three subsamples overlap strongly and exhibit a broad range of luminosities, including both detections and upper limits. Below $M_{\ast}{\sim}10^{8}\,\mathrm{M_{\odot}}$, we see almost exclusively upper limits.

Fig.~\ref{pl_lum_sfr} shows the relation between the X-ray luminosity and SFR for the galaxies in our late-type sample. The left panel displays the 0.5--2\,keV luminosity of the thermal component, while the right panel shows the total 0.5--8\,keV luminosity. Detections and upper limits are treated separately in the fit, and the blue lines indicate the best-fit relations with their corresponding $\pm1\sigma$ intrinsic scatter. The black and magenta lines show the comparison relations from \citet{mineo2}, \citet{mineo3}, and \cite{kyritsis2025}. The green shaded regions show the prediction from the X-ray luminosity function (XLF)-based model of \citet{kyritsis2026} for the integrated luminosity of unresolved X-ray binaries, including the scatter expected from stochastic sampling of the XRB population. Olive green indicates the 99\,\% range, dark green the 99.9\,\% range.

The black data point in the left panel shows a theoretical prediction for the thermal X-ray luminosity from simulations (\citealt{rodgers-lee2019}; Coombs et al. in prep) that follow superbubble evolution in an isolated disk galaxy on a 3D whole-galaxy scale. For this comparison, the simulation was run with a feedback efficiency of 50\,\%, so that half of the mechanical energy from stellar winds and supernovae is dissipated locally, while the remainder drives superbubble expansion and outflows into the halo. This choice is observationally motivated by starburst-driven winds in M82 \citep{strickland2009}. The 0.5--2\,keV luminosity was obtained by assigning a \texttt{mekal} spectrum to each simulation cell and integrating over the band following \citet{krause2014}; the error bar gives the $1\sigma$ temporal variation over 110\,Myr.

For the total integrated luminosity in 0.5--8\,keV, the best-fit relation is
\begin{equation}\label{eq_tot}
    \log L_\mathrm{X,tot} = (39.5 \pm 0.4) + (0.87 \pm 0.20)\,\log \mathrm{SFR},
\end{equation}
with an intrinsic scatter of $\sigma_\mathrm{int} = (0.87\pm0.18)\,\mathrm{dex}$. For the thermal component in 0.5--2\,keV, we find
\begin{equation}\label{eq_apec}
    \log L_\mathrm{X,apec} = (38.7 \pm 0.4) + (1.07 \pm 0.24)\,\log \mathrm{SFR},
\end{equation}
with $\sigma_\mathrm{int} = (0.73\pm0.17)\,\mathrm{dex}$ at 90\,\% confidence. The thermal $L_\mathrm{X}-\mathrm{SFR}$ relation is steeper and has a smaller intrinsic scatter than the total relation.

\begin{table*}
\centering
\caption{Fit results and absorption-corrected fluxes for the stacked spectra.}             
\label{tab_res3}      
\begin{tabular}{lccccc}      
\hline\hline                  
        Parameter                    & Unit                                      & High                   & Medium                  & Low                       & ETG                          \\\hline
        sSFR                         & $10^{-10}\,\mathrm{yr}^{-1}$              & $\geq 2.0$             & $\geq 0.5$              & $< 0.5$                   &                              \\ 
        Sample size                  &                                           & 22                     & 47$^{\ast}$             & 57                        & 13                           \\\vspace{5pt}
        Cts                          &                                           & 1598                   & 1612                    & 619                       & 688                         \\
        cstat/dof                    &                                           & 1.08                   & 1.07                    & 1.06                      & 1.06                          \\
        $N_\mathrm{H,Gal}$           & $10^{22}\,\mathrm{cm^{-2}}$               & 0.029                  & 0.039                   & 0.071                     & 0.038                       \\
        $N_\mathrm{H,int}$           & $10^{22}\,\mathrm{cm^{-2}}$               & $0.24^{+0.11}_{-0.09}$ & $0.20^{+0.17}_{-0.11}$  & $0.28^{+0.33\,+0.26}_{-0.19\,-0.02}$ & $0.33^{+0.32}_{-0.26}$         \\
        $kT$                         & keV                                       & $0.76^{+0.13}_{-0.17}$ & $0.82^{+0.08}_{-0.09}$  & $0.2,\,0.3,\,0.5$         & $0.26^{+0.35}_{-0.11}$           \vspace{2pt}\\
        $\text{norm}_\text{apec}$    &   $10^{-4}\,\mathrm{cm^{-5}}$             & $1.6^{+0.7}_{-0.7}$    & $1.5^{+0.9}_{-0.5}$     & $<\,1.5\,\,(<\,9.0)$        & $3.4^{+26}_{-1.8}$                  \vspace{2pt}\\
        $\Gamma$                     &                                           & $2.1^{+0.6}_{-0.5}$    & $<\,2.2$                & $1.8^{+0.7\,+0.2}_{-0.7}$ & $3.5^{+1.8}_{-1.4}$                         \vspace{2pt}\\
        $\text{norm}_\text{PL}$      & $10^{-4}\,\mathrm{cts\,keV^{-1}\,cm^{-2}\,s^{-1}}$
                                                                                 & $5.5^{+2.0}_{-1.5}$    & $1.7^{+1.0}_{-0.5}$     & $0.9^{+0.5\,+0.5}_{-0.3\,-0.1}$      & $4.4^{+4.9}_{-2.1}$   \vspace{10pt}\\
        $F_\text{apec}^{\mathrm{0.5\mbox{--}2\,keV}}$  & $10^{-14}\,\mathrm{erg\,s^{-1}\,cm^{-2}}$
                                                                                 & $5.4^{+2.3}_{-1.9}$    & $5.2^{+3.0}_{-1.6}$     & $<\,2.9\,\,(<\,5.8)$         & $7.4^{+17}_{-5.9}$ \vspace{2pt}\\
        $F_\text{tot}^{\mathrm{0.5\mbox{--}2\,keV}}$   & $10^{-13}\,\mathrm{erg\,s^{-1}\,cm^{-2}}$     
                                                                                 & $2.8^{+0.7}_{-0.5}$    & $1.2^{+0.5}_{-0.3}$     & $0.5^{+0.3\,+0.1}_{-0.2}$ & $3.3^{+7.1}_{-1.8}$   \vspace{2pt}\\
        $F_\text{tot}^{\mathrm{0.5\mbox{--}8\,keV}}$   & $10^{-13}\,\mathrm{erg\,s^{-1}\,cm^{-2}}$     
                                                                                 & $4.8^{+0.6}_{-0.5}$    & $3.0^{+1.0}_{-0.8}$     & $1.0^{+0.4}_{-0.2}$       & $3.3^{+3.7}_{-1.3}$    \vspace{2pt}\\
\hline                  
\end{tabular}
\begin{tablenotes}
\footnotesize
\item \textit{Notes.} Early-Type (ETG) and high-, medium-, and low-sSFR stacked spectra. The sSFR is calculated using the SFRs and stellar masses from HECATEv2 \citep{hecatev2}. Fit parameters are similar to Table~\ref{tab_res1}. For the low-sSFR spectrum, we explored a grid of fixed plasma temperature values $kT = 0.2, 0.3, 0.5\,\mathrm{keV}$. We use $kT=0.3\,\mathrm{keV}$ for the reference model shown in Fig.~\ref{spec_stacked}. Where two uncertainties are given, the first denotes the statistical $90\,\%$ confidence interval of the reference model, and the second the systematic range, defined by the lowest lower bound and highest upper bound among the $90\,\%$ confidence intervals of all tested models. The bracketed flux values correspond to the most conservative upper limit given by the $kT=0.2\,\mathrm{keV}$ model. $^{\ast}$NGC 5236 was excluded from the stacked spectrum.
\end{tablenotes}
    \end{table*}

\section{Discussion}\label{chap5}

\subsection{Spectra}

The individually analyzed galaxies demonstrate that the relative contributions of thermal gas and compact sources vary strongly between systems. This motivated us to treat the thermal and total X-ray emission separately. In the following, we focus on the behavior of these components in the stacked galaxy populations.

The stacked spectra show that both the thermal and nonthermal flux decrease toward lower sSFR. At the same time, the thermal component becomes weaker relative to the nonthermal continuum in the low-sSFR stack compared to the high- and medium-sSFR stacks. This is consistent with the decline of stellar feedback at low star-forming activity: as the energy input from massive stars and supernovae decreases, the emission from hot gas becomes weaker, while X-ray emission from older compact-source populations remains present. The high-sSFR stack additionally contains ULXs, which strongly increase its nonthermal flux and show how individual bright sources can affect the integrated spectrum.

The plasma temperature also points to a change in the dominant thermal component. The high- and medium-sSFR spectra are described by a relatively hot plasma with $kT {\sim} 0.8\,\mathrm{keV}$, while in the low-sSFR spectrum the thermal component is too weak to be constrained and is consistent with assumed values of $kT = 0.2\mbox{--}0.5\,\mathrm{keV}$. This can also be explained by hot-phase gas fading with declining star-forming activity and leading to cooler and weaker thermal emission.

The early-type galaxies show a different behavior. Although star formation is weak or absent, a clear thermal component is still detected at a relatively cool temperature of $kT {\sim} 0.26\,\mathrm{keV}$. This suggests that hot gas can remain present even without ongoing stellar feedback, likely because it is retained in the deeper gravitational potential of these galaxies. At the same time, the best-fit continuum is considerably softer, with $\Gamma \sim 3.5$ compared to $\Gamma \sim 2$ for the star-forming subsamples. This is consistent with a reduced contribution from the hard HMXB population associated with recent star formation and a larger contribution from older LMXB populations.

\subsection{Virialized halo gas}

A possible source of soft thermal X-ray emission is hot gas in the galactic halo. Gas accreted onto sufficiently massive galaxies can be shock-heated to the virial temperature and form an extended, quasi-hydrostatic halo \citep[e.g.,][]{white1991}. In this scenario, the temperature of the X-ray emitting gas $T_\mathrm{vir}$ is linked to the gravitational potential of the dark matter halo $M_\mathrm{halo}$ and scales with $T_\mathrm{vir} \propto M_\mathrm{halo}^{2/3}$ \citep[e.g.,][]{barkana2001}. This leads to a plasma temperature of $kT {\sim} 0.2\,\mathrm{keV}$ for the Milky Way \citep[e.g.,][]{miller2015}. Since stellar mass correlates with halo mass \citep{moster2013, behroozi2013}, a thermal component dominated by virialized halo gas would be expected to become hotter in more massive galaxies.

For the medium-sSFR spectrum, we find a plasma temperature of $kT {\sim} 0.82\,\mathrm{keV}$, which is far above the expected virial temperature for Milky-Way-like halos. The individually fit galaxies also show no clear temperature-mass trend: the relatively massive galaxies NGC~5236 and NGC~1291 show moderate effective plasma temperatures of $kT{\sim}0.3\mbox{--}0.4\,\mathrm{keV}$, while the less massive galaxy NGC~1313 has a higher best-fit temperature of $kT {\sim} 0.88\,\mathrm{keV}$. While we cannot rule out a contribution from virialized halo gas, the fitted temperatures indicate that the thermal component is unlikely to be primarily driven by the gravitational potential. Instead, they favor an origin in stellar-feedback-heated gas. This interpretation is also supported by the superbubble simulation shown in Fig.~\ref{pl_lum_sfr}, where the predicted 0.5--2\,keV luminosity agrees with our observed $L_\mathrm{X}-\mathrm{SFR}$ relation.

\subsection{$L_\mathrm{X}-\mathrm{SFR}$ relation}

The main result of this work is that the total X-ray luminosity and the luminosity of the thermal component follow different scaling relations with SFR. The thermal relation remains consistent with a linear scaling over the sampled range from SFRs of ${\sim}5\,\mathrm{M_\odot\,yr^{-1}}$ down to ${\sim}10^{-5}\,\mathrm{M_\odot\,yr^{-1}}$, supporting the interpretation that thermal X-ray emission is a direct tracer of stellar feedback-heated gas in the ISM. In contrast, the total relation is flatter and exhibits a larger intrinsic scatter, indicating that the two quantities are influenced by partly different physical processes.

The normalization $A$ in Eq.~\ref{eq_log_fit} gives the logarithmic X-ray luminosity predicted at $\mathrm{SFR}=1\,\mathrm{M_\odot\,yr^{-1}}$. For the thermal component, our normalization $A=38.73\pm0.32$ corresponds to $L_\mathrm{X}=(5.4^{+5.7}_{-2.8})\cdot10^{38}\,\mathrm{erg\,s^{-1}}$, which is in good agreement with the 0.5--2\,keV value of $L_\mathrm{X}=(5.2\pm0.2)\cdot10^{38}\,\mathrm{erg\,s^{-1}}$ derived by \citet{mineo2}. Likewise, the normalization $A=39.46\pm0.32$ of the total relation corresponds to $L_\mathrm{X}=(2.9^{+3.1}_{-1.5})\cdot10^{39}\,\mathrm{erg\,s^{-1}}$, which is close to the 0.5--8\,keV value of $L_\mathrm{X}=(4.0\pm0.4)\cdot10^{39}\,\mathrm{erg\,s^{-1}}$ obtained by \citet{mineo3}. By comparison, \citet{kyritsis2025} find a higher logarithmic normalization of ${\sim}40.05$ in 0.5--8\,keV, which they associate with a population of X-ray-luminous galaxies, particularly at low SFR. In addition, \citet{mineo2,mineo3} show that the fitted normalization can be sensitive to the adopted fitting procedure and SFR calibration.

\begin{figure*}[t!]
    \centering
        \includegraphics[width=.47\textwidth]{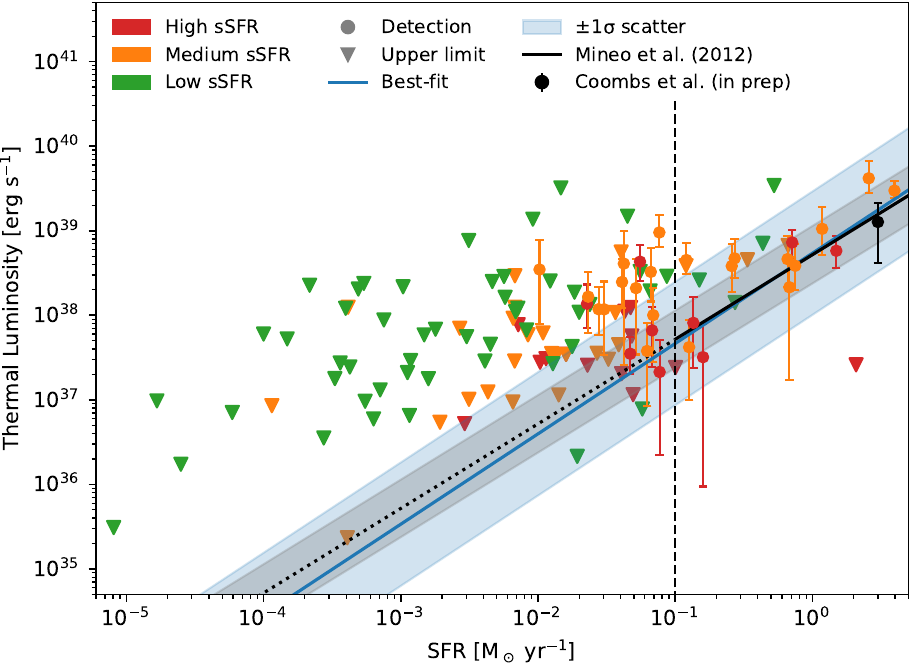}
        \includegraphics[width=.47\textwidth]{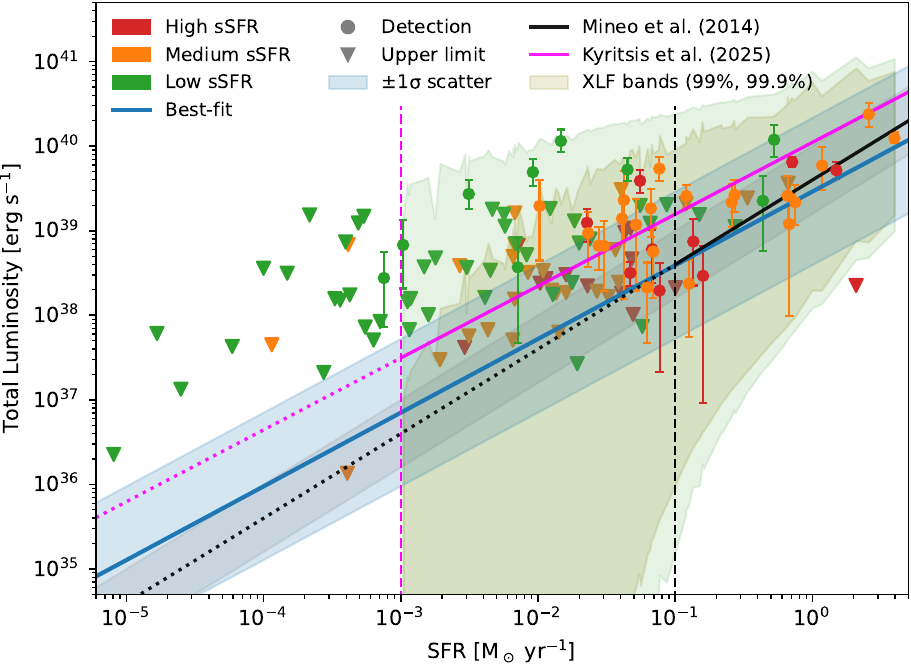} 
         \caption{0.5--2\,keV luminosity of the thermal component over the SFR (left) and integrated 0.5--8\,keV luminosity over the SFR (right). High-, medium-, and low-sSFR galaxies are shown in red, orange, and green. Circles mark detections, triangles mark upper limits. The blue lines show the best-fit relations, with the shaded regions indicating the $\pm1\sigma$ intrinsic scatter. In the left panel, the black line shows the comparison relation from \citet{mineo2}; the black data point shows a theoretical prediction based on \citet{rodgers-lee2019} and Coombs et al. (in prep). In the right panel, the black and magenta lines show the comparison relations from \citet{mineo3} and \citet{kyritsis2025}, respectively. The dashed vertical lines mark the lower SFR threshold of the samples used to derive these relations. Below these thresholds, the comparison relations are shown as dotted-line extrapolations for comparison only. The green shaded regions mark the XLF-based luminosity range enclosing $99\,\%$ and $99.9\,\%$ of the expected integrated point-source luminosity from \citet{kyritsis2026}.}
         \label{pl_lum_sfr}
\end{figure*}

\begin{figure}[b!]
    \centering
    \includegraphics[width=.94\linewidth]{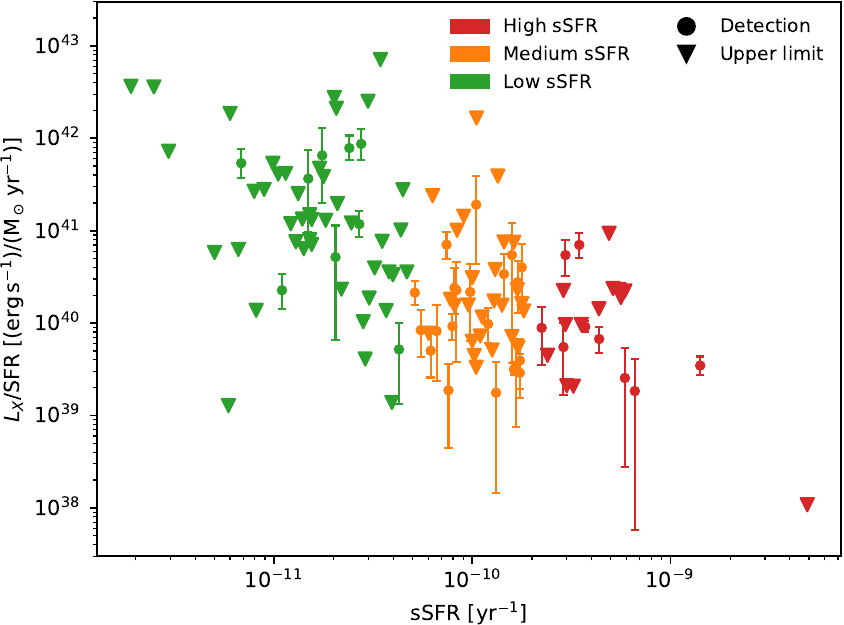}
    \caption{Total 0.5--8\,keV luminosity per unit SFR, $L_\mathrm{X}/\mathrm{SFR}$, as a function of sSFR, for the star-forming galaxies with reliable SFRs. High-, medium-, and low-sSFR galaxies are shown in red, orange, and green. Circles mark detections, triangles mark upper limits.}
    \label{lxsfrssfr}
\end{figure}

The total $L_\mathrm{X}-\mathrm{SFR}$ relation is flatter and exhibits a larger intrinsic scatter. This indicates that, in a blind sample, the integrated X-ray luminosity is no longer primarily shaped by the hot ISM. Instead, the total luminosity becomes increasingly sensitive to nonthermal contributions from unresolved XRBs. This interpretation is consistent with the fit results, which show that the nonthermal continuum contribution grows toward the low-SFR regime. The right panel in Fig.~\ref{pl_lum_sfr} shows that many low-sSFR galaxies have upper limits well above the best-fit relation, and some are detected at an elevated luminosity compared to it. The comparison with the X-ray luminosity function (XLF) bands from \citet{kyritsis2026} suggests that these detections can be understood as stochastic outliers, produced when the XRB population is no longer sampled smoothly, but is dominated by small-number statistics. This interpretation is further supported by the distribution of $L_\mathrm{X}/\mathrm{SFR}$ values as a function of sSFR (Fig.~\ref{lxsfrssfr}), which shows a visible upturn and a much larger scatter toward lower sSFR. This indicates that some low-sSFR galaxies are too X-ray luminous to be explained by stellar feedback alone, consistent with their X-ray output being boosted by a small number of bright XRBs. Stochastic sampling of the compact-source population may therefore contribute to the flattening of the total $L_\mathrm{X}-\mathrm{SFR}$ relation.

The observed scatter may, however, also be affected by uncertainties in the adopted SFR estimates. As discussed in Sect.~\ref{sect_catalogs}, the HECATEv2 SFRs are homogenized from different indicators, which may produce systematic differences. We examined the origin of the adopted SFR estimates across the three sSFR subsamples and three distance bins (see Table~\ref{sfr_origin_table} in Appendix~\ref{appendix_sfr}). The high- and medium-sSFR subsamples are dominated by L19 UV+WISE estimates, whereas WISE-only and K23 estimates become more common toward lower sSFR. No strong dependence on distance is observed. We also verified that the pileup around $\log(\mathrm{sSFR/yr^{-1}})\sim-11.5$, reported for the L19 calibration, does not occur in our sample. 

Variations in the metallicity may also provide an additional contribution to the scatter. The median metallicities of the high-, medium-, and low-sSFR subsamples are $12+\log(\mathrm{O/H})=8.29$, 8.39, and 8.38, respectively. The widths of the corresponding 16th--84th percentile intervals increase from 0.26, to 0.44, and 0.52\,dex toward lower sSFR. There is thus no systematic shift in metallicity between the three sSFR subsamples, but the increasing metallicity spread toward lower sSFR may contribute to the $L_\mathrm{X}$ scatter at a given SFR. The corresponding metallicity distributions are shown in Fig.~\ref{appendix_metallicity} in Appendix~\ref{appendix_metal}.

We also tested a two-component relation of the form
\begin{equation}\label{2d_fit_eq}
    \log L_\mathrm{X} = \log\left(\alpha\mathrm{SFR}+\beta M_\ast\right)
\end{equation}
and found that, although the fit converges, the two coefficients $\alpha$ and $\beta$ are strongly degenerate and cannot be independently constrained with our current sample. The two-component fit is shown in Fig.~\ref{appendix_2dfit}; the likelihood contours of both the two-component, and the SFR-only fit, are shown in Fig.~\ref{appendix_contours_fig} in Appendix~\ref{appendix_contours}.

\subsection{Comparison to previous studies}

Our eROSITA sample includes 140 galaxies within 25\,Mpc and spans six orders of magnitude in SFR, down to $\sim10^{-5}\,\mathrm{M_\odot\,yr^{-1}}$, with sSFRs of $10^{-12} \mbox{--} 10^{-9}\,\mathrm{yr^{-1}}$ for the star-forming subsample. This reaches well below the ranges covered by most previous studies and extends significantly below the SFR range studied by \citet{mineo2}. In contrast to pointed studies that are naturally biased toward bright and well-detected systems, our sample includes galaxies from the full local population, including very faint systems. A comparison with the samples of \citet{grimes2005}, \citet{owen2009}, and \citet{li2013} also shows that only four of the 127 star-forming galaxies in our sample overlap with these studies of hot gas X-ray emission.

For the thermal $L_\mathrm{X}-\mathrm{SFR}$ relation, our results are in agreement with \citet{mineo2}, who investigated the diffuse X-ray emission of 29 star-forming galaxies ($\mathrm{SFR} {\sim} 0.1 \mbox{--} 17\,\mathrm{M_\odot\,yr^{-1}}$) with \textit{Chandra}, and found that the diffuse 0.5--2\,keV emission of nearby star-forming galaxies scales linearly with the SFR. Our best-fit thermal slope of $\alpha = 1.07 \pm 0.24$ remains consistent with a linear scaling within uncertainties. Due to the broader parameter range covered by our sample, and stochastic variations at lower SFRs, our intrinsic scatter of $(0.73\pm0.17)\,\mathrm{dex}$ is larger than the $\sim0.34\,\mathrm{dex}$ quoted by \citet{mineo2}. 

For the total luminosity in 0.5--8\,keV, \citet{mineo3} also derived a linear scaling relation ($\sigma_\mathrm{int} {\sim} 0.4\,\mathrm{dex}$), using a \textit{Chandra} sample of 66 galaxies with $\mathrm{SFR} {\sim} 0.1 \mbox{--} 10^{3}\,\mathrm{M_\odot\,yr^{-1}}$. With a slope of $\alpha = 0.87 \pm 0.20$, and an intrinsic scatter of $(0.87\pm0.18)\,\mathrm{dex}$, our relation is both flatter and broader. In the low SFR regime studied in our sample, the integrated X-ray output appears to flatten, consistent with increasing contributions from compact source populations that are not directly related to the current rate of star-formation.

This interpretation is consistent with the results presented by \citet{gilfanov2004letter} and \citet{gilfanov2004}. These studies showed that the low-SFR regime is affected by (a) a strongly asymmetric luminosity distribution caused by small-number statistics and (b) an increasing contribution from LMXBs that scales with stellar mass. Together, these effects can explain the upward spread that we see in both the total $L_\mathrm{X}-\mathrm{SFR}$ (Fig.~\ref{pl_lum_sfr}, right), and the $L_\mathrm{X}/\mathrm{SFR}$ (Fig.~\ref{lxsfrssfr}) diagrams. This is also consistent with \citet{lehmer2010}, who found that both SFR and stellar mass are required in galaxy $L_\mathrm{X}$ scaling relations.

The closest blind-survey comparison is \citet{kyritsis2025}. Using a sample of 18790 star-forming galaxies in eRASS:1, they derive a slope of $0.85^{+0.03}_{-0.09}$, which is in good agreement with our result for the integrated X-ray luminosity. \citet{kyritsis2025} also found that integrated X-ray luminosities are elevated and broadly distributed, especially at low SFR, and argue that stochastic XRB sampling alone is not sufficient to explain the observed excess; stellar population age and metallicity must also play a role. Our results are consistent with this picture: bright XRBs likely explain part of the flattening and upward spread, but an additional dependence on stellar mass, age, or metallicity remains plausible. A similar conclusion was reached by \citet{lehmer2019}, who found that galaxy X-ray scaling relations depend on stellar population age and metallicity in addition to SFR. This motivates a follow-up study with a larger sample, aimed at exploring the multidimensional relation between $L_\mathrm{X}$, SFR, stellar mass, metallicity, and age.

\section{Conclusion}\label{chap6}

We analyzed the integrated X-ray emission of a volume-limited, blind sample of 140 nearby galaxies within 25\,Mpc using eRASS:4 data. By combining the Updated Nearby Galaxy Catalog \citep{karachentsev} with HECATEv2 \citep{hecatev2}, we constructed a statistically complete sample of local galaxies, spanning six orders of magnitude in SFR, down to ${\sim}10^{-5}\,\mathrm{M_\odot\,yr^{-1}}$. This allows us to probe the hot ISM in a parameter range that is largely inaccessible to pointed X-ray studies, and to derive scaling relations that are not limited to bright, star-forming systems.

Our stacked spectra reveal a systematic change in the origin of the integrated X-ray emission along the sSFR sequence. High-sSFR galaxies show the strongest soft thermal component, consistent with a large contribution from stellar-feedback-heated gas. There is also no clear evidence for a correlation of the temperature of the thermal component with stellar mass. The measured temperatures are generally higher than the virial temperatures expected for the galaxies in our sample. This further strengthens the conclusion that the thermal component is related to star-formation-related processes, such as superbubbles, rather than being primarily set by the depth of the gravitational potential. Toward lower sSFR, the spectra become increasingly dominated by the power-law continuum. In the low-sSFR stack, the thermal component is only marginally constrained. Early-type galaxies still show a cool thermal component, possibly gas that is retained in the deep gravitational potential, but the spectrum suggests a dominant contribution from older compact-source populations.

The main result of this work is that the thermal and total X-ray luminosities yield different best-fit scaling relations with SFR. The thermal component shows a steep relation with a slope of $\alpha = 1.07 \pm 0.24$ and an intrinsic scatter of $(0.73 \pm 0.17)\,\mathrm{dex}$, whereas the total luminosity follows a flatter relation with $\alpha = 0.87 \pm 0.20$ and a larger scatter of $(0.87 \pm 0.18)\,\mathrm{dex}$ at 90\,\% confidence. The thermal $L_\mathrm{X}-\mathrm{SFR}$ relation remains close to linear over the SFR range covered by our sample, which suggests that the thermal soft X-ray emission stays linked to stellar feedback and the hot ISM, even in very weakly star-forming galaxies. This is further supported by the agreement between the prediction from superbubble evolution in an isolated galaxy simulation and our observational results, as demonstrated in Fig.~\ref{pl_lum_sfr}. Meanwhile, the total luminosity is no longer a clean tracer of star-formation-related processes in the low-SFR regime.

A plausible explanation is that, in a blind galaxy sample, the integrated X-ray output becomes increasingly sensitive to unresolved XRBs when the SFR declines. The upward scatter in total luminosity, the increase in $L_\mathrm{X}/\mathrm{SFR}$ toward low SFR, and the agreement with the XLF-based expectations from \citet{kyritsis2026} indicate that the total X-ray emission is affected by stochastic compact-source emission, and possibly by an increasing dependence on stellar mass and stellar population properties. A two-component SFR- and stellar-mass-dependent fit is strongly degenerate, such that the individual contributions cannot be robustly constrained with the present sample; residual SFR-estimator and metallicity variations may additionally contribute to the observed scatter.

In conclusion, our results show that the thermal soft X-ray component is the more reliable tracer for studies of the hot ISM in nearby galaxies. The total X-ray luminosity becomes increasingly difficult to interpret in the low-SFR regime, where compact sources can dominate the signal. eROSITA now makes it possible to see this transition in a statistically complete local sample. This gives future studies an observational basis for exploring how galaxy X-ray emission depends on SFR, stellar mass, metallicity, stellar population age and the contribution of XRBs.

\bibliographystyle{aa}
{\footnotesize
\bibliography{references}}

\begin{acknowledgements}

This work was supported by the Deutsche Forschungsgemeinschaft (DFG) through the project
SA 2131/18-1 (project no. 525647185). S.H. is partly supported by the German Science Foundation (DFG grant numbers WI 1860/14-1 and 434448349). MGHK acknowledges financial support by the DFG through the project SA 2131/15-2 (project no. 414059771) as a Mercator Fellow in the Research Unit FOR2990 eRO-STEP. The paper is based on data from eROSITA, the soft X-ray instrument aboard SRG, a joint Russian-German science mission supported by
the Russian Space Agency (Roskosmos), in the interests of the Russian Academy
of Sciences represented by its Space Research Institute (IKI), and the Deutsches
Zentrum für Luft- und Raumfahrt (DLR). The SRG spacecraft was built by Lavochkin Association (NPOL) and its subcontractors and is operated by NPOL
with support from the Max-Planck Institute for Extraterrestrial Physics (MPE).
The development and construction of the eROSITA X-ray instrument were led
by MPE, with contributions from the Dr. Karl Remeis Observatory Bamberg,
the Erlangen Center for Astroparticle Physics (ECAP), the University of Hamburg Observatory, the Leibniz Institute for Astrophysics Potsdam (AIP), and the
Institute for Astronomy and Astrophysics of the University of Tübingen, with
the support of DLR and the Max Planck Society. The Argelander Institute for
Astronomy of the University of Bonn and the Ludwig Maximilians Universität
Munich also participated in the science preparation for eROSITA. The eROSITA
data shown here were processed using the \texttt{eSASS/NRTA} software system developed by the German eROSITA consortium.

\end{acknowledgements}

\clearpage
\begin{appendix}
    
\onecolumn
\setlength{\parindent}{0pt}

\section{Galaxy sample properties}\label{appendix_metal}

\begin{figure*}[hbp!]
    \centering
    \begin{tabular}{cc}
        \includegraphics[height=0.26\textheight]{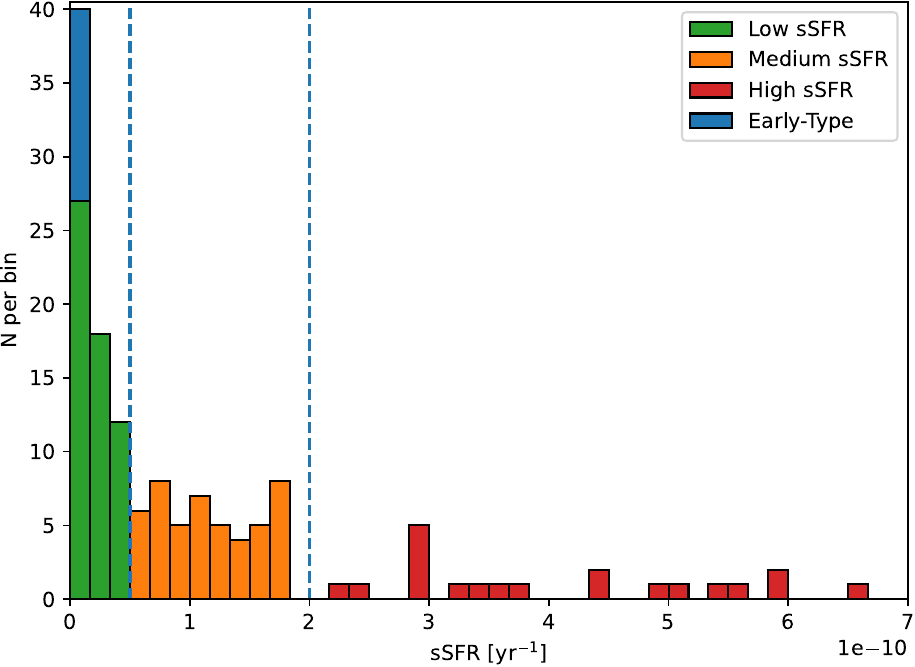} & 
        \includegraphics[height=0.26\textheight]{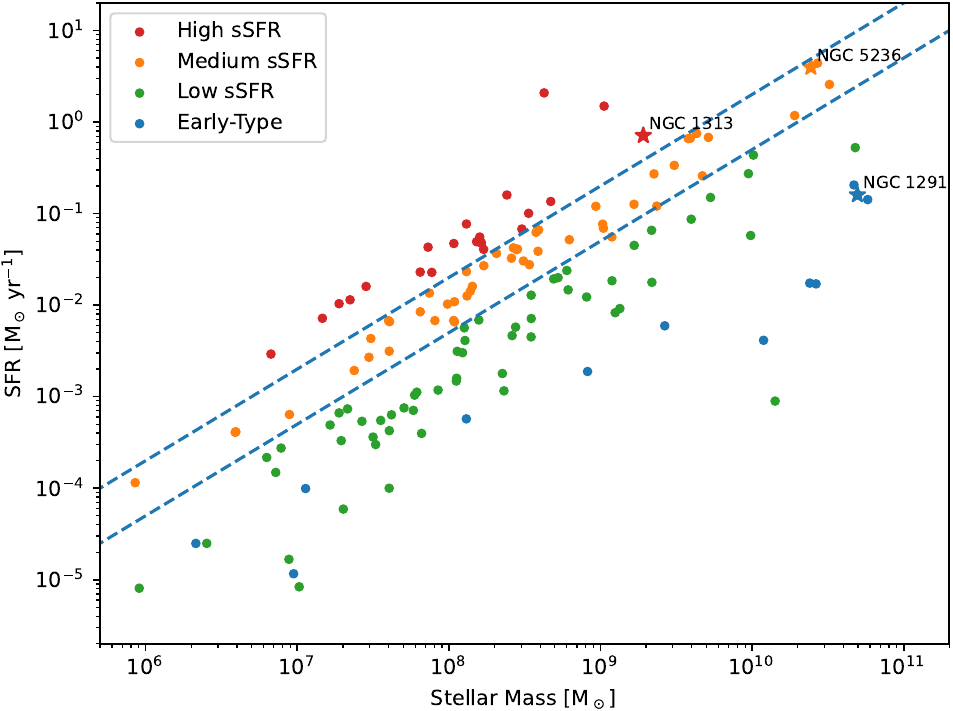}\\
    \end{tabular}
      \caption{Distribution of sSFRs (left) and SFR over stellar mass (right) for our galaxy sample. High-, medium-, and low-sSFR galaxies are shown in red, orange, and green, while early-type galaxies are shown in blue. The dashed lines represent the sSFR thresholds of $0.5\cdot10^{-10}$ and $2.0\cdot10^{-10}\,\mathrm{yr^{-1}}$. The three individually analyzed galaxies NGC~5236, NGC~1291, and NGC~1313 are marked by star symbols. SFRs and stellar masses are taken from HECATEv2 \citep{hecatev2}.}
         \label{mass_sfr}
\end{figure*}

In this appendix, we summarize the properties of the galaxy sample. Fig.~\ref{mass_sfr} shows the distribution of sSFR and SFR as a function of stellar mass for the high-, medium-, and low-sSFR subsamples and the early-type galaxies.

We additionally show the gas-phase metallicity distributions of the galaxies used in the $L_\mathrm{X}-\mathrm{SFR}$ relation. HECATEv2 metallicity estimates are available for 120 out of the 122 galaxies (21 high-, 47 medium, and 52 low-sSFR galaxies). The metallicity is expressed in units of $12+\log(\mathrm{O/H})$. We obtain

\begin{align*}
\mathrm{High\ sSFR:}\quad
12+\log(\mathrm{O/H}) &= 8.29^{+0.15}_{-0.11}\;(\Delta=0.26~\mathrm{dex}),\\
\mathrm{Medium\ sSFR:}\quad
12+\log(\mathrm{O/H}) &= 8.39^{+0.25}_{-0.18}\;(\Delta=0.44~\mathrm{dex}),\\
\mathrm{Low\ sSFR:}\quad
12+\log(\mathrm{O/H}) &= 8.38^{+0.24}_{-0.27}\;(\Delta=0.52~\mathrm{dex}).
\end{align*}

\begin{figure}[thbp!]
    \centering
    \includegraphics[width=0.5\linewidth]{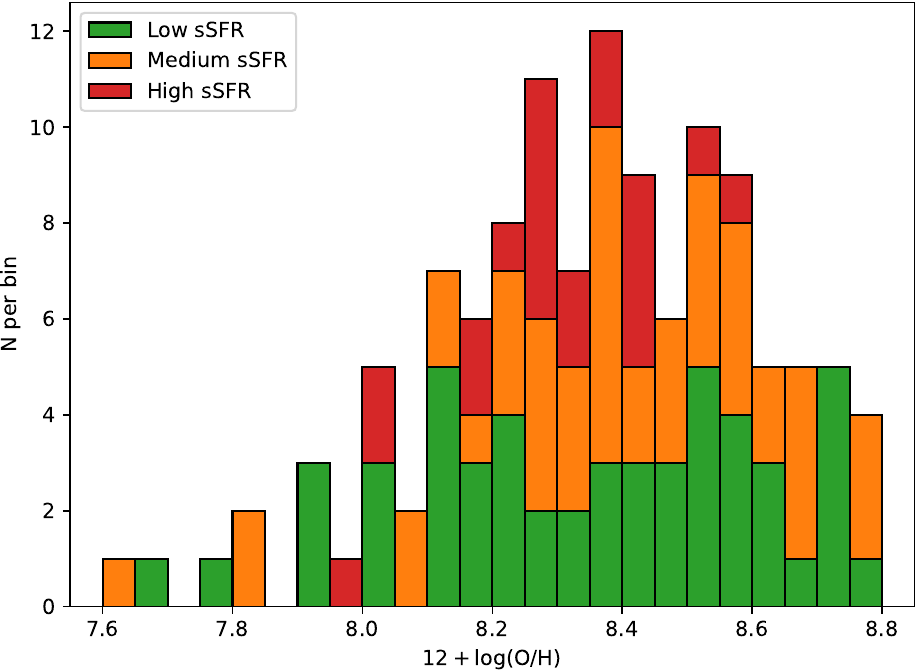}
    \caption{Distribution of the gas-phase metallicities, $12+\log(\mathrm{O/H})$, for the high-, medium-, and low-sSFR galaxies used in the $L_\mathrm{X}-\mathrm{SFR}$ analysis. Only galaxies with available HECATEv2 metallicity estimates are shown.}
    \label{appendix_metallicity}
\end{figure}

The metallicity values denote the median, the asymmetric uncertainties give the 16th--84th percentile ranges, and $\Delta$ denotes the width of these intervals. The corresponding distributions are shown in Fig.~\ref{appendix_metallicity}.

\clearpage

\section{Spectral fits}\label{appendix_spectra}

\begin{figure}[htbp!]
    \centering
    \includegraphics[width=.5\textwidth]{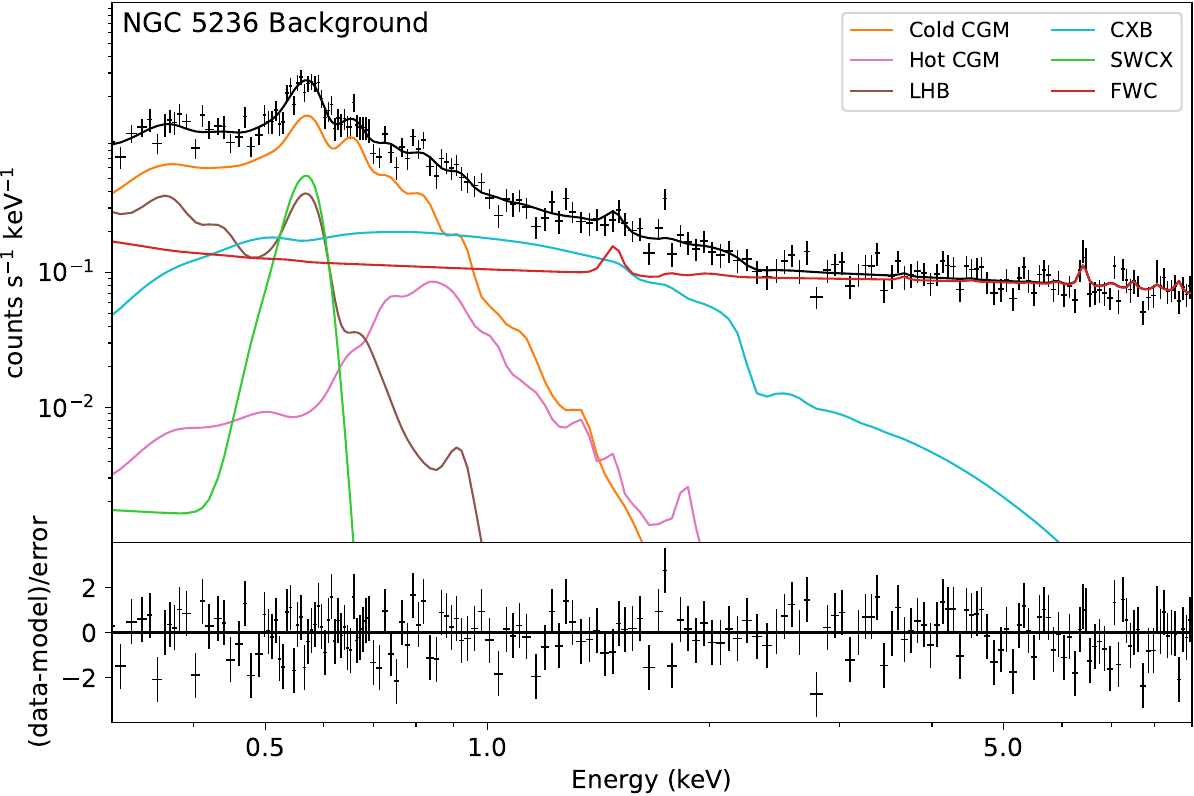}  
      \caption{eRASS:4 background spectrum of NGC 5236. The black line shows the full spectral model, the colored lines show the contributions of the different model components.}
         \label{bkg_spectrum}
\end{figure}

\begin{figure}[htbp!]
    \centering
    \begin{tabular}{cc}
        \includegraphics[height=0.24\textheight]{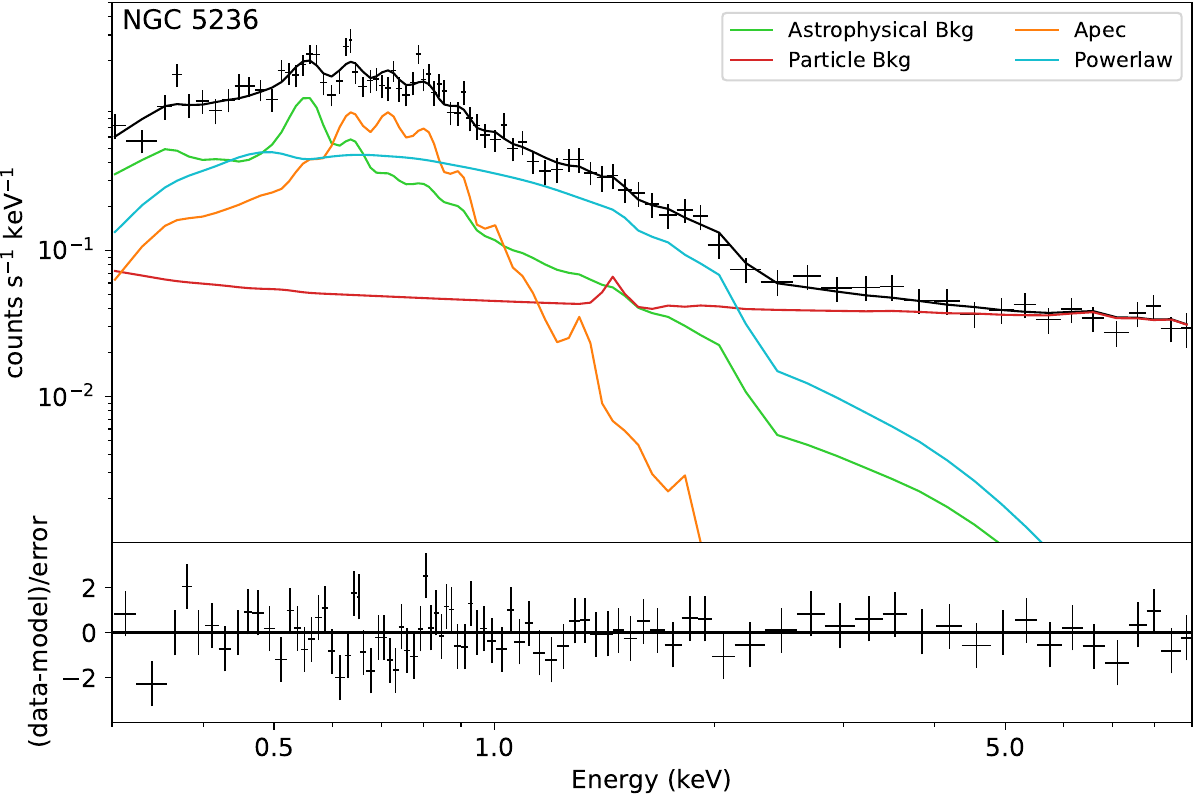} &
        \includegraphics[height=0.24\textheight]{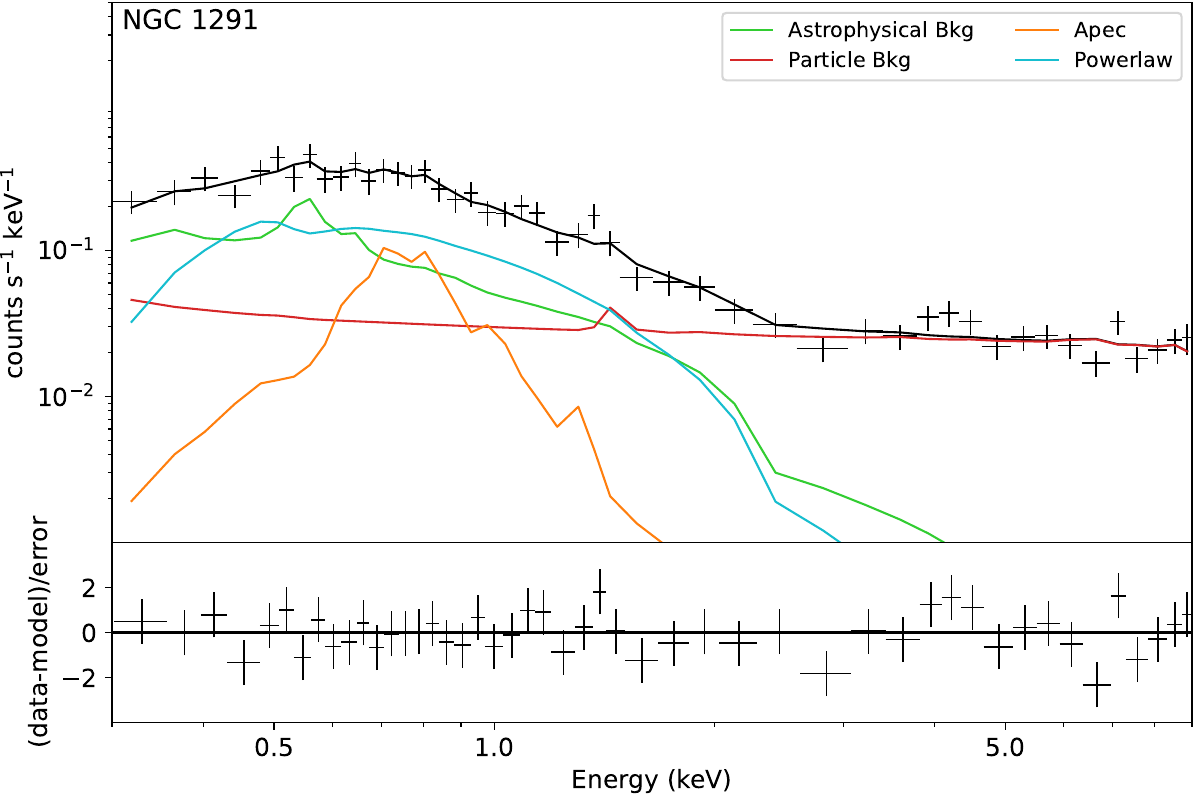} \\  
        \includegraphics[height=0.24\textheight]{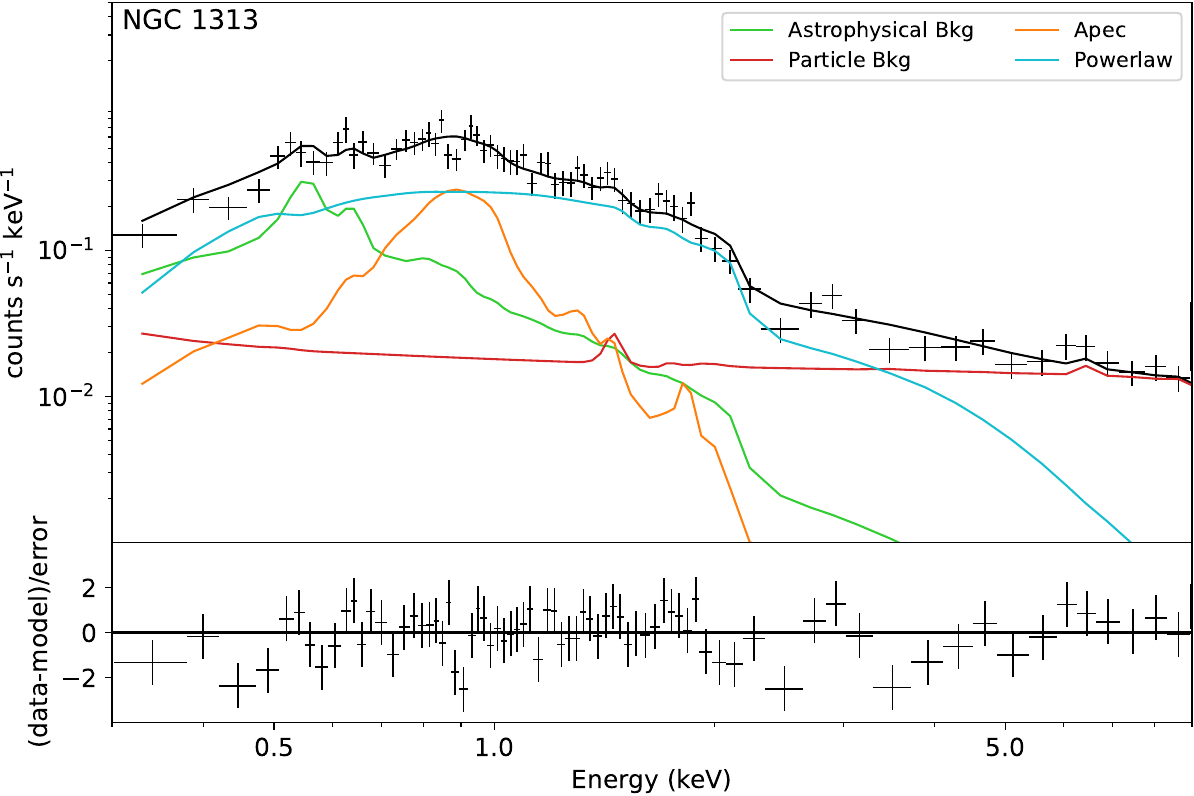} &
        \includegraphics[height=0.24\textheight]{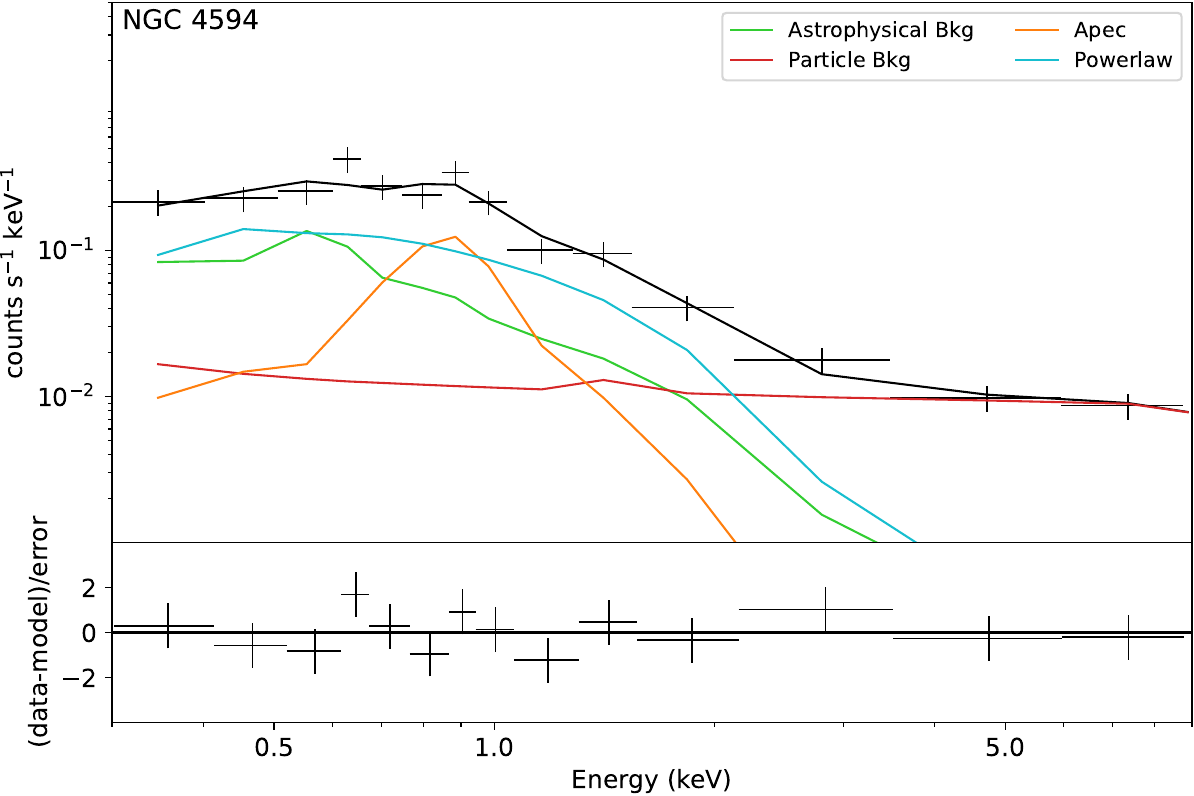} \\
    \end{tabular}
      \caption{eRASS:4 spectra of NGC~5236 (upper left), NGC~1291 (upper right), NGC~1313 (lower left), and NGC~4594 (lower right). The black lines show the full spectral model, the colored lines show the contributions of the different model components.}
         \label{spectra}
\end{figure}

This appendix shows the spectral fits used in the analysis. Fig.~\ref{bkg_spectrum} illustrates the decomposition of the background model, while Fig.~\ref{spectra} shows the source spectra and best-fit components of the individually fitted galaxies.

\newpage

\section{SFR systematics}\label{appendix_sfr}

In this appendix, we show the origin of the adopted HECATEv2 estimates for the galaxies used in the $L_\mathrm{X}-\mathrm{SFR}$ analysis. We compare their relative contributions across the three sSFR subsamples and three distance bins. The results are shown in Table~\ref{sfr_origin_table}.

\begin{table}[htbp!]
\centering
\caption{Origin of the adopted HECATEv2 SFR estimates for the galaxies used in the $L_\mathrm{X}-\mathrm{SFR}$ analysis, separated by sSFR and distance.}
\label{sfr_origin_table}
\begin{tabular}{lrrrrr}
\hline
Sample/Distance & $N$ & L19 UV+WISE & L19 WISE4 & K23 & S18 SED \\
\hline
High-sSFR       & 22 & 18 (81.8\,\%) &  2 (9.1\,\%)  &  2 (9.1\,\%)  & 0 (0\,\%) \\
Medium-sSFR     & 47 & 37 (78.7\,\%) &  4 (8.5\,\%)  &  6 (12.8\,\%) & 0 (0\,\%) \\
Low-sSFR        & 53 & 16 (30.2\,\%) & 14 (26.4\,\%) & 23 (43.4\,\%) & 0 (0\,\%) \\
\hline
$< 5\,\mathrm{Mpc}$         & 20 & 13 (65.0\,\%) &  2 (10.0\,\%) &  5 (25.0\,\%) & 0 (0\,\%) \\
$5\mbox{--}10\,\mathrm{Mpc}$ & 70 & 41 (58.6\,\%) & 14 (20.0\,\%) & 15 (21.4\,\%) & 0 (0\,\%) \\
$\geq 10\,\mathrm{Mpc}$     & 32 & 17 (53.1\,\%) &  4 (12.5\,\%) & 11 (34.4\,\%) & 0 (0\,\%) \\
\hline
\end{tabular}
\begin{tablenotes}
\footnotesize
\item \textit{Notes.} Percentages are given relative to the total number of galaxies in each subsample. L19 UV+WISE denotes the hybrid FUV+WISE4 or NUV+WISE4 SFR estimates from Leroy et al. (2019), while L19 WISE4 denotes their WISE4-only estimates. K23 refers to the infrared/optical SFR calibrations of Kouroumpatzakis et al. (2023), and S18 SED to the UV-to-optical spectral energy distribution fits from Salim et al. (2018).
\end{tablenotes}
\end{table}

\section{Likelihood function for the $L_\mathrm{X}-\mathrm{SFR}$ fits}\label{appendix_like}

Here we describe the maximum-likelihood approach used to fit the $L_\mathrm{X}-\mathrm{SFR}$ relations in (Eq.~\ref{eq_apec}) and (Eq.~\ref{eq_tot}). The fit was performed in logarithmic space, with $x_\mathrm{i} = \log\mathrm{SFR_i}$ and $z_\mathrm{i} = \log L_\mathrm{X,i}$. Detections were modeled as Gaussian measurements around the relation, while upper limits were treated as censored data. Asymmetric uncertainties in luminosity were symmetrized in log-space assuming Gaussian measurement errors in $\log L_\mathrm{X}$
\begin{equation}\label{eq_symmetrize}
    \sigma_{\log L} = \tfrac{1}{2}\left[\log_{10}(L+\Delta L_+) - \log_{10}(L-\Delta L_-)\right].
\end{equation}
For detections, the likelihood contribution of each object in the likelihood function is 
\begin{equation}
    \mathcal{L}_\mathrm{det,i} \propto
    \frac{1}{\sqrt{2\pi(\sigma_\mathrm{int}^2 + \sigma_\mathrm{obs,i}^2)}} 
    \exp\!\left[-\frac{(z_\mathrm{i} - A - \alpha x_\mathrm{i})^2}
    {2(\sigma_\mathrm{int}^2 + \sigma_\mathrm{obs,i}^2)}\right],
\end{equation}
where $\sigma_\mathrm{obs, i}$ is the measurement error, and $\sigma_\mathrm{int}$ the intrinsic scatter. For upper limits, the data do not provide an exact luminosity measurement, but the condition that the value lies below $z_\mathrm{lim,i}$. Their likelihood is therefore given by the integrated Gaussian probability below this threshold, i.e., by the cumulative normal distribution function $\Phi(u)$:
\begin{equation}
    \mathcal{L}_\mathrm{lim,i} = 
    \Phi\!\left(\frac{z_\mathrm{lim,i} - A - \alpha x_\mathrm{i}}
    {\sigma_\mathrm{int}}\right), \;\;\;\;\; \Phi(u) = \frac{1}{\sqrt{2\pi}} \int\limits_{-\infty}^{u} e^{-t^2/2}\, \mathrm{d}t
\end{equation}
with the cumulative normal distribution function $\Phi$ and the upper limit expressed in units of the intrinsic scatter relative to the model prediction $u_\mathrm{i} = (z_\mathrm{lim,i} - A - \alpha x_\mathrm{i})/\sigma_\mathrm{int}$. The total negative log-likelihood is then
\begin{equation}\label{neglog}
    -\ln \mathcal{L} = 
    \frac{1}{2} \sum_\mathrm{det}
    \left[\ln(2\pi s_\mathrm{i}^2) + 
    \frac{(z_\mathrm{i} - A - \alpha x_\mathrm{i})^2}{s_\mathrm{i}^2}\right]
    - \sum_\mathrm{lim} \ln \Phi(u_\mathrm{i}),
\end{equation}
with $s_\mathrm{i}^2 = \sigma_\mathrm{int}^2 + \sigma_{\mathrm{obs,i}}^2$.
The fit parameters $A$, $\alpha$, and $\sigma_\mathrm{int}$ were determined by minimizing the negative log-likelihood in Eq. \ref{neglog} using \texttt{scipy.optimize.minimize} with the BFGS algorithm in \texttt{Python} (version 3.11.14). The uncertainties are 90\,\% confidence errors.

\section{Two-component fit and confidence contours}\label{appendix_contours}

\begin{figure}[htbp!]
    \centering
    \includegraphics[width=0.5\linewidth]{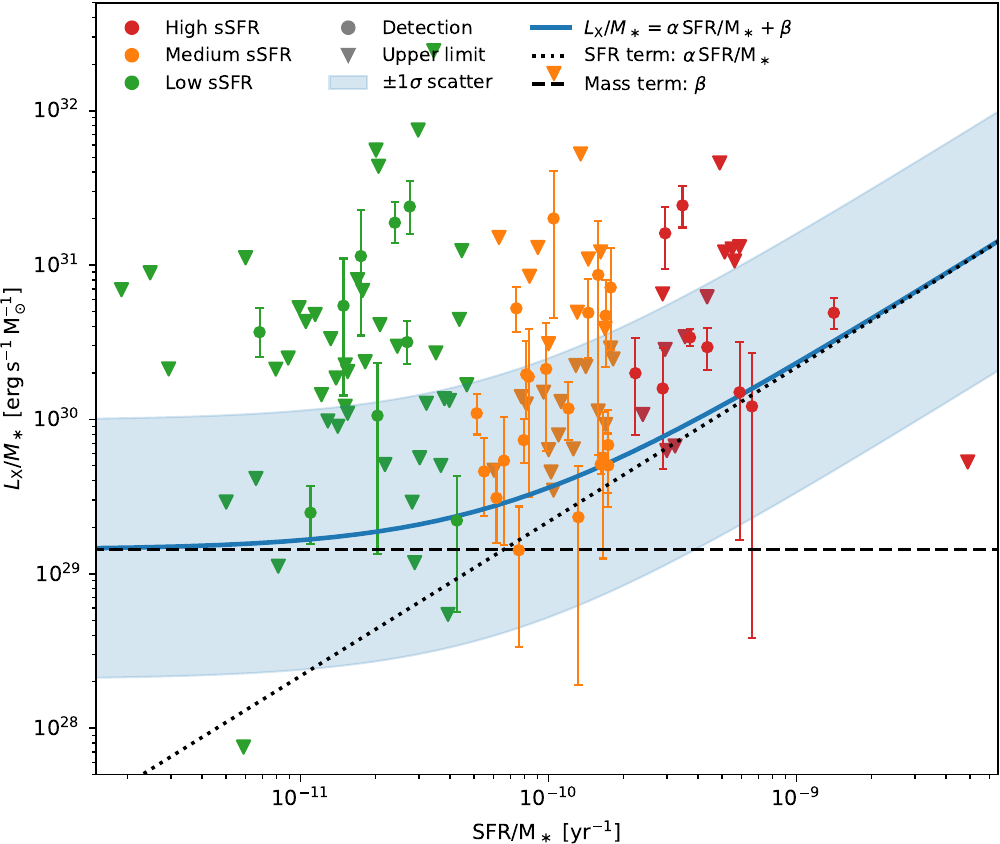}
    \caption{Two-component fit to the total X-ray luminosity in 0.5--8\,keV, shown in units of $L_\mathrm{X}/M_\ast$ as a function of $\mathrm{SFR}/M_\ast$. The blue line shows the best-fit relation, the dotted and dashed lines show the individual SFR and stellar-mass terms. The blue region indicates the $1\sigma$ intrinsic scatter. High-, medium-, and low-sSFR galaxies are shown in red, orange, and green; circles indicate detections and triangles upper limits.}
    \label{appendix_2dfit}
\end{figure}

In addition to the $L_\mathrm{X}-\mathrm{SFR}$ relation, we fitted the integrated X-ray luminosity in 0.5--8\,keV with the two-component model shown in Eq.~\ref{2d_fit_eq}. We used the same censored maximum-likelihood approach as described in Appendix~\ref{appendix_like}, and performed the fit for the same 122 galaxies used for the $L_\mathrm{X}-\mathrm{SFR}$ relation. The fit is shown in Fig.~\ref{appendix_2dfit}.
We obtain
\begin{align*}
    \log \alpha &= 39.3^{+0.4}_{-0.7},\\
    \log \beta  &= 29.2^{+0.4}_{-1.0},\\
    \sigma_{\mathrm{int}} &= 0.84^{+0.19}_{-0.14}\,\mathrm{dex},
\end{align*}
where $\alpha$ is given in $(\mathrm{erg\,s^{-1}})/(\mathrm{M_\odot\,yr^{-1}})$ and $\beta$ in $\mathrm{erg\,s^{-1}\,M_\odot}^{-1}$. The quoted uncertainties are the 90\,\% confidence intervals.

\begin{figure}[htbp!]
    \centering
    \begin{tabular}{cc}
        \includegraphics[height=0.3\textheight]{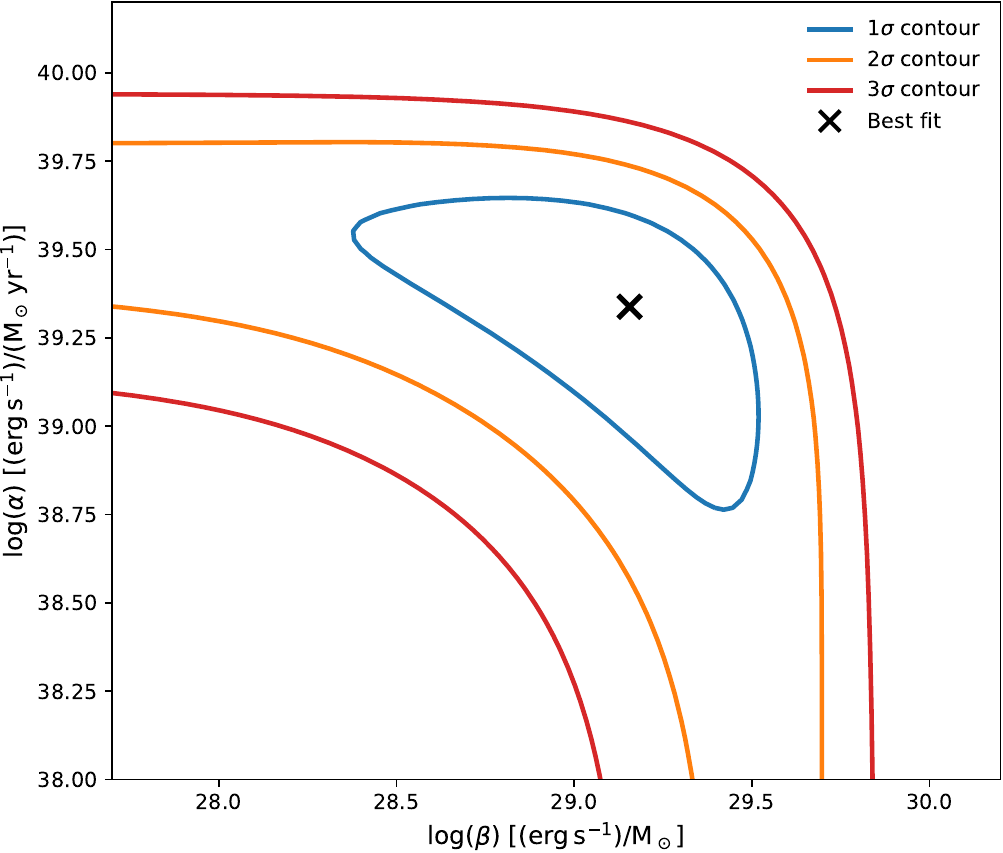} &
        \includegraphics[height=0.3\textheight]{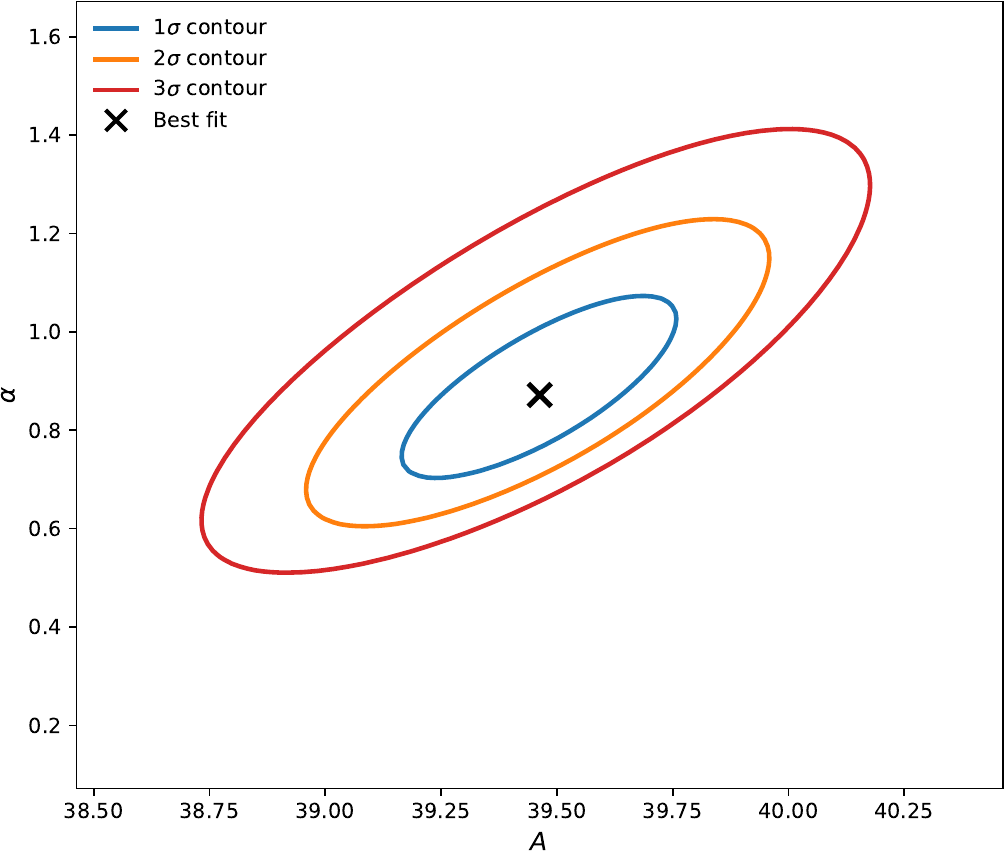} \\  
    \end{tabular}
    \caption{Confidence contours for the $L_\mathrm{X}$ scaling relation fits. Left: joint confidence contours of the SFR coefficient $\alpha$ and the stellar-mass coefficient $\beta$ from the two-component relation in Eq.~\ref{2d_fit_eq}. Right: joint confidence contours of the normalization $A$ and the slope $\alpha$ from the SFR-only relation in Eq.~\ref{eq_log_fit}. The blue, orange, and red curves show the $1\sigma$, $2\sigma$, and $3\sigma$ contours; the black crosses mark the best-fit values.}
    \label{appendix_contours_fig}
\end{figure}

We also calculated confidence contours for the fitted parameters. For the two-component fit, we varied $\log\alpha$ and $\log\beta$ around their best-fit values and calculated $\sigma_\mathrm{int}$ at each point. The resulting contours are shown in the left panel in Fig.~\ref{appendix_contours_fig}. For comparison, we also show the confidence contours for the normalization and slope of the SFR-only fit in the right panel in Fig.~\ref{appendix_contours_fig}.

\clearpage

\section{Galaxy properties}\label{appendix_prop}

This appendix lists the galaxy properties and the derived X-ray luminosities. Table~\ref{tab_high} lists the properties of the high-sSFR galaxies, Table~\ref{tab_med} the medium-sSFR galaxies, Table~\ref{tab_low1} the low-sSFR galaxies, and Table~\ref{tab_etg} the early-type galaxies.

\begin{table}[htbp!]
\caption{Properties of high sSFR galaxy sample.}
\label{tab_high}
\centering
\footnotesize
\resizebox{\textwidth}{!}{%
\begin{tabular}{lccccccc}
\hline\hline
Name & Dist. & Cts & SFR & $M_\ast$ & sSFR & $L_{\mathrm{X,tot}}^{\mathrm{0.5-8\,keV}}$ & $L_{\mathrm{X,th}}^{\mathrm{0.5-2\,keV}}$ \\
 & (Mpc) &  & ($\mathrm{M_\odot\,yr}^{-1}$) & ($\mathrm{M_\odot}$) & ($\mathrm{yr}^{-1}$) & (\si{erg.s^{-1}}) & (\si{erg.s^{-1}}) \\
\hline
UGCA193     & 9.7  & 0.3    & $2.08\cdot10^{0}$  & $4.25\cdot10^{8}$ & $4.90\cdot10^{-9}$  & $<2.2\cdot10^{38}$              & $<2.6\cdot10^{37}$              \\
ESO495-021  & 8.0  & 76.1   & $1.49\cdot10^{0}$  & $1.06\cdot10^{9}$ & $1.41\cdot10^{-9}$  & $5.2^{+1.3}_{-1.1}\cdot10^{39}$ & $5.8^{+2.8}_{-2.2}\cdot10^{38}$ \\
ESO572-034  & 8.5  & 4.3    & $1.60\cdot10^{-1}$ & $2.41\cdot10^{8}$ & $6.61\cdot10^{-10}$ & $3.0^{+3.5}_{-2.9}\cdot10^{38}$ & $3.2^{+4.6}_{-3.1}\cdot10^{37}$ \\
DDO068      & 9.8  & 3.3    & $4.29\cdot10^{-2}$ & $7.29\cdot10^{7}$ & $5.89\cdot10^{-10}$ & $<9.6\cdot10^{38}$              & $<1.1\cdot10^{38}$              \\
NGC1592     & 9.1  & 4.3    & $7.69\cdot10^{-2}$ & $1.31\cdot10^{8}$ & $5.89\cdot10^{-10}$ & $1.9^{+2.2}_{-1.7}\cdot10^{38}$ & $2.1^{+2.9}_{-1.9}\cdot10^{37}$ \\
DDO064      & 7.1  & 1.3    & $1.60\cdot10^{-2}$ & $2.84\cdot10^{7}$ & $5.62\cdot10^{-10}$ & $<3.0\cdot10^{38}$              & $<3.5\cdot10^{37}$              \\
ESO149-003  & 5.9  & 3.4    & $1.04\cdot10^{-2}$ & $1.89\cdot10^{7}$ & $5.50\cdot10^{-10}$ & $1.2^{+1.5}_{-1.2}\cdot10^{38}$ & $1.3^{+2.0}_{-1.3}\cdot10^{37}$ \\
IC3023      & 7.7  & 1.1    & $1.14\cdot10^{-2}$ & $2.23\cdot10^{7}$ & $5.13\cdot10^{-10}$ & $<2.7\cdot10^{38}$              & $<3.1\cdot10^{37}$              \\
DDO083      & 10.9 & 0.9    & $7.18\cdot10^{-3}$ & $1.47\cdot10^{7}$ & $4.90\cdot10^{-10}$ & $<6.7\cdot10^{38}$              & $<7.7\cdot10^{37}$              \\
NGC1705     & 5.1  & 56.5   & $4.71\cdot10^{-2}$ & $1.08\cdot10^{8}$ & $4.37\cdot10^{-10}$ & $3.2^{+1.1}_{-0.9}\cdot10^{38}$ & $3.5^{+1.9}_{-1.4}\cdot10^{37}$ \\
MRK0036     & 8.0  & $-0.2$ & $2.92\cdot10^{-3}$ & $6.71\cdot10^{6}$ & $4.36\cdot10^{-10}$ & $<4.2\cdot10^{37}$              & $<5.3\cdot10^{36}$              \\
NGC1313     & 4.1  & 1353.0 & $7.12\cdot10^{-1}$ & $1.92\cdot10^{9}$ & $3.72\cdot10^{-10}$ & $6.5^{+0.9}_{-0.8}\cdot10^{39}$ & $7.3^{+3.1}_{-2.5}\cdot10^{38}$ \\
ESO245-005  & 4.4  & 8.2    & $2.30\cdot10^{-2}$ & $6.47\cdot10^{7}$ & $3.55\cdot10^{-10}$ & $<2.3\cdot10^{38}$              & $<2.6\cdot10^{37}$              \\
NGC4765     & 9.7  & 42.3   & $5.54\cdot10^{-2}$ & $1.60\cdot10^{8}$ & $3.47\cdot10^{-10}$ & $3.9^{+1.3}_{-1.1}\cdot10^{39}$ & $4.3^{+2.4}_{-1.8}\cdot10^{38}$ \\
UGC05456    & 5.6  & 0.1    & $4.93\cdot10^{-2}$ & $1.52\cdot10^{8}$ & $3.24\cdot10^{-10}$ & $<1.0\cdot10^{38}$              & $<1.1\cdot10^{37}$              \\
UGCA298     & 11.0 & 0.2    & $1.01\cdot10^{-1}$ & $3.36\cdot10^{8}$ & $3.00\cdot10^{-10}$ & $<2.1\cdot10^{38}$              & $<2.4\cdot10^{37}$              \\
IC3591      & 9.8  & $-3.5$ & $4.82\cdot10^{-2}$ & $1.63\cdot10^{8}$ & $2.95\cdot10^{-10}$ & $<4.6\cdot10^{38}$              & $<5.7\cdot10^{37}$              \\
ESO486-021  & 10.0 & 20.7   & $2.28\cdot10^{-2}$ & $7.72\cdot10^{7}$ & $2.95\cdot10^{-10}$ & $1.2^{+0.6}_{-0.5}\cdot10^{39}$ & $1.4^{+0.9}_{-0.7}\cdot10^{38}$ \\
NGC3274     & 8.0  & 8.7    & $1.36\cdot10^{-1}$ & $4.70\cdot10^{8}$ & $2.88\cdot10^{-10}$ & $7.5^{+6.1}_{-5.2}\cdot10^{38}$ & $8.2^{+8.3}_{-5.8}\cdot10^{37}$ \\
DDO084      & 8.1  & 5.0    & $4.73\cdot10^{-2}$ & $1.64\cdot10^{8}$ & $2.88\cdot10^{-10}$ & $<1.1\cdot10^{39}$              & $<1.2\cdot10^{38}$              \\
VCC1725     & 8.5  & $-0.2$ & $4.07\cdot10^{-2}$ & $1.70\cdot10^{8}$ & $2.40\cdot10^{-10}$ & $<1.8\cdot10^{38}$              & $<2.0\cdot10^{37}$              \\
NGC4700     & 7.3  & 12.3   & $6.79\cdot10^{-2}$ & $3.03\cdot10^{8}$ & $2.24\cdot10^{-10}$ & $6.0^{+4.2}_{-3.5}\cdot10^{38}$ & $6.6^{+6.0}_{-4.1}\cdot10^{37}$ \\
\hline
\end{tabular}%
}
\tablefoot{Columns list the galaxy name, distance, net counts, star formation rate, stellar mass, specific star formation rate, total luminosity, and (\texttt{apec}) soft-band luminosity of the thermal component. For non-detections, 90\,\% upper limits are given.}
\end{table}

\begin{table}[htbp!]
\caption{Properties of medium sSFR galaxy sample.}
\label{tab_med}
\centering
\footnotesize
\resizebox{\textwidth}{!}{%
\begin{tabular}{lccccccc}
\hline\hline
Name & Dist. & Cts & SFR & $M_\ast$ & sSFR & $L_{\mathrm{X,tot}}^{\mathrm{0.5-8\,keV}}$ & $L_{\mathrm{X,th}}^{\mathrm{0.5-2\,keV}}$ \\
 & (Mpc) &  & ($\mathrm{M_\odot\,yr}^{-1}$) & ($\mathrm{M_\odot}$) & ($\mathrm{yr}^{-1}$) & (\si{erg.s^{-1}}) & (\si{erg.s^{-1}}) \\
\hline
ESO347-017    & 7.6  & 0.0    & $1.36\cdot10^{-2}$ & $7.45\cdot10^{7}$  & $1.82\cdot10^{-10}$ & $<1.8\cdot10^{38}$               & $<3.4\cdot10^{37}$               \\
DDO161        & 7.3  & 19.3   & $2.32\cdot10^{-2}$ & $1.31\cdot10^{8}$  & $1.78\cdot10^{-10}$ & $9.3^{+7.4}_{-5.6}\cdot10^{38}$  & $1.6^{+1.6}_{-1.0}\cdot10^{38}$  \\
NGC4455       & 8.4  & 1.5    & $3.67\cdot10^{-2}$ & $2.06\cdot10^{8}$  & $1.78\cdot10^{-10}$ & $<5.9\cdot10^{38}$               & $<1.1\cdot10^{38}$               \\
UGCA127       & 8.5  & 26.0   & $7.49\cdot10^{-1}$ & $4.31\cdot10^{9}$  & $1.74\cdot10^{-10}$ & $2.2^{+1.3}_{-1.0}\cdot10^{39}$  & $3.8^{+3.1}_{-1.9}\cdot10^{38}$  \\
NGC3432       & 9.2  & 20.5   & $6.59\cdot10^{-1}$ & $3.79\cdot10^{9}$  & $1.74\cdot10^{-10}$ & $2.6^{+1.8}_{-1.3}\cdot10^{39}$  & $4.6^{+3.9}_{-2.4}\cdot10^{38}$  \\
NGC2835       & 10.3 & 12.4   & $6.65\cdot10^{-1}$ & $3.92\cdot10^{9}$  & $1.70\cdot10^{-10}$ & $<3.6\cdot10^{39}$               & $<6.6\cdot10^{38}$               \\
NGC4204       & 8.0  & 23.2   & $6.64\cdot10^{-2}$ & $3.91\cdot10^{8}$  & $1.70\cdot10^{-10}$ & $1.8^{+1.3}_{-1.0}\cdot10^{39}$  & $3.3^{+2.9}_{-1.8}\cdot10^{38}$  \\
LV-J1235-1104 & 10.0 & $-0.7$ & $6.76\cdot10^{-3}$ & $4.00\cdot10^{7}$  & $1.69\cdot10^{-10}$ & $<1.6\cdot10^{38}$               & $<2.9\cdot10^{37}$               \\
NGC0625       & 3.9  & 17.9   & $6.24\cdot10^{-2}$ & $3.76\cdot10^{8}$  & $1.66\cdot10^{-10}$ & $2.1^{+2.0}_{-1.7}\cdot10^{38}$  & $3.8^{+4.2}_{-2.9}\cdot10^{37}$  \\
NGC5236       & 4.9  & 1309   & $3.96\cdot10^{0}$  & $2.44\cdot10^{10}$ & $1.62\cdot10^{-10}$ & $1.2^{+0.3}_{-0.2}\cdot10^{40}$ & $3.0^{+1.0}_{-0.6}\cdot10^{39}$ \\
UGC05427      & 7.1  & 2.8    & $6.59\cdot10^{-3}$ & $4.06\cdot10^{7}$  & $1.62\cdot10^{-10}$ & $<4.9\cdot10^{38}$               & $<9.1\cdot10^{37}$               \\
IC1959        & 6.0  & 5.1    & $2.69\cdot10^{-2}$ & $1.70\cdot10^{8}$  & $1.58\cdot10^{-10}$ & $<1.9\cdot10^{38}$               & $<3.6\cdot10^{37}$               \\
UGC04704      & 15.2 & 5.5    & $4.22\cdot10^{-2}$ & $2.66\cdot10^{8}$  & $1.58\cdot10^{-10}$ & $2.3^{+2.9}_{-2.1}\cdot10^{39}$  & $4.1^{+5.8}_{-3.8}\cdot10^{38}$  \\
NGC3104       & 16.0 & 2.1    & $4.04\cdot10^{-2}$ & $2.80\cdot10^{8}$  & $1.45\cdot10^{-10}$ & $<3.1\cdot10^{39}$               & $<5.6\cdot10^{38}$               \\
\end{tabular}
}
\end{table}

\begin{table}[htbp!]
\label{tab_med2}
\centering
\footnotesize
\resizebox{\textwidth}{!}{%
\begin{tabular}{lccccccc}

IC3583        & 7.6  & 22.3   & $4.10\cdot10^{-2}$ & $2.84\cdot10^{8}$  & $1.45\cdot10^{-10}$ & $1.4^{+0.9}_{-0.7}\cdot10^{39}$  & $2.5^{+2.1}_{-1.3}\cdot10^{38}$  \\
IC4247        & 5.0  & 0.4    & $4.32\cdot10^{-3}$ & $3.06\cdot10^{7}$  & $1.41\cdot10^{-10}$ & $<6.8\cdot10^{37}$               & $<1.2\cdot10^{37}$               \\
HIPASS-J1337-39 & 4.8 & $-1.4$ & $1.10\cdot10^{-4}$ & $8.53\cdot10^{5}$ & $1.35\cdot10^{-10}$ & $<4.5\cdot10^{37}$              & $<8.6\cdot10^{36}$               \\
NGC1637       & 9.3  & 14.8   & $6.80\cdot10^{-1}$ & $5.16\cdot10^{9}$  & $1.32\cdot10^{-10}$ & $1.2^{+1.4}_{-1.1}\cdot10^{39}$  & $2.1^{+2.8}_{-1.9}\cdot10^{38}$  \\
VCC0841       & 7.9  & 1.3    & $8.46\cdot10^{-3}$ & $6.48\cdot10^{7}$  & $1.31\cdot10^{-10}$ & $<3.2\cdot10^{38}$               & $<5.9\cdot10^{37}$               \\
NGC4597       & 10.1 & 7.8    & $1.20\cdot10^{-1}$ & $9.32\cdot10^{8}$  & $1.29\cdot10^{-10}$ & $<2.1\cdot10^{39}$               & $<3.9\cdot10^{38}$               \\
NGC5264       & 4.5  & 1.1    & $3.26\cdot10^{-2}$ & $2.59\cdot10^{8}$  & $1.26\cdot10^{-10}$ & $<1.7\cdot10^{38}$               & $<3.1\cdot10^{37}$               \\
NGC5068       & 5.4  & 113.0  & $2.71\cdot10^{-1}$ & $2.26\cdot10^{9}$  & $1.20\cdot10^{-10}$ & $2.6^{+1.3}_{-1.0}\cdot10^{39}$  & $4.7^{+3.3}_{-1.9}\cdot10^{38}$  \\
NGC1311       & 5.2  & 6.0    & $1.60\cdot10^{-2}$ & $1.43\cdot10^{8}$  & $1.12\cdot10^{-10}$ & $<1.9\cdot10^{38}$               & $<3.4\cdot10^{37}$               \\
IC5332        & 7.8  & 13.9   & $3.36\cdot10^{-1}$ & $3.07\cdot10^{9}$  & $1.10\cdot10^{-10}$ & $<2.4\cdot10^{39}$               & $<4.6\cdot10^{38}$               \\
ESO499-038    & 15.3 & 0.4    & $4.10\cdot10^{-4}$ & $3.94\cdot10^{6}$  & $1.05\cdot10^{-10}$ & $<6.8\cdot10^{38}$               & $<1.3\cdot10^{38}$               \\
LeoA          & 0.8  & $-10.5$ & $4.10\cdot10^{-4}$ & $3.90\cdot10^{6}$ & $1.05\cdot10^{-10}$ & $<1.2\cdot10^{36}$              & $<2.1\cdot10^{35}$               \\
DDO097        & 13.7 & 6.7    & $1.03\cdot10^{-2}$ & $9.80\cdot10^{7}$  & $1.05\cdot10^{-10}$ & $2.0^{+2.0}_{-1.5}\cdot10^{39}$  & $3.5^{+4.2}_{-2.6}\cdot10^{38}$  \\
IC5152        & 2.0  & 0.1    & $1.42\cdot10^{-2}$ & $1.39\cdot10^{8}$  & $1.02\cdot10^{-10}$ & $<6.1\cdot10^{37}$               & $<1.1\cdot10^{37}$               \\
DDO131        & 8.1  & 1.5    & $1.09\cdot10^{-2}$ & $1.09\cdot10^{8}$  & $1.00\cdot10^{-10}$ & $<3.3\cdot10^{38}$               & $<6.1\cdot10^{37}$               \\
NGC1800       & 8.0  & 1.5    & $3.87\cdot10^{-2}$ & $3.87\cdot10^{8}$  & $1.00\cdot10^{-10}$ & $<2.5\cdot10^{38}$               & $<4.6\cdot10^{37}$               \\
NGC3037       & 8.1  & 7.6    & $3.03\cdot10^{-2}$ & $3.10\cdot10^{8}$  & $9.80\cdot10^{-11}$ & $6.7^{+6.4}_{-4.8}\cdot10^{38}$  & $1.2^{+1.3}_{-0.9}\cdot10^{38}$  \\
UGC06014      & 10.4 & $-0.5$ & $1.26\cdot10^{-2}$ & $1.32\cdot10^{8}$ & $9.50\cdot10^{-11}$ & $<2.0\cdot10^{38}$              & $<3.5\cdot10^{37}$               \\
FS04          & 10.4 & 0.4    & $2.69\cdot10^{-3}$ & $2.97\cdot10^{7}$  & $9.10\cdot10^{-11}$ & $<3.8\cdot10^{38}$               & $<7.0\cdot10^{37}$               \\
IC2992        & 12.7 & 0.6    & $6.78\cdot10^{-3}$ & $8.09\cdot10^{7}$  & $8.40\cdot10^{-11}$ & $<6.7\cdot10^{38}$               & $<1.2\cdot10^{38}$               \\
ESO300-014    & 9.8  & 25.0   & $5.18\cdot10^{-2}$ & $6.23\cdot10^{8}$  & $8.30\cdot10^{-11}$ & $1.2^{+1.2}_{-1.0}\cdot10^{39}$  & $2.1^{+2.5}_{-1.7}\cdot10^{38}$  \\
ESO483-013    & 7.4  & 15.5   & $2.76\cdot10^{-2}$ & $3.40\cdot10^{8}$  & $8.10\cdot10^{-11}$ & $6.7^{+4.3}_{-3.2}\cdot10^{38}$  & $1.2^{+1.0}_{-0.6}\cdot10^{38}$  \\
SexB          & 1.4  & 1.0    & $1.93\cdot10^{-3}$ & $2.38\cdot10^{7}$  & $8.10\cdot10^{-11}$ & $<3.0\cdot10^{37}$               & $<5.5\cdot10^{36}$               \\
NGC2903       & 8.9  & 205.5  & $2.57\cdot10^{0}$  & $3.24\cdot10^{10}$ & $7.90\cdot10^{-11}$ & $2.4^{+0.9}_{-0.7}\cdot10^{40}$  & $4.2^{+2.5}_{-1.4}\cdot10^{39}$  \\
UGCA319       & 7.3  & $-1.1$ & $3.14\cdot10^{-3}$ & $4.04\cdot10^{7}$ & $7.80\cdot10^{-11}$ & $<5.7\cdot10^{37}$              & $<1.0\cdot10^{37}$               \\
NGC0300       & 2.2  & 82.9   & $1.26\cdot10^{-1}$ & $1.66\cdot10^{9}$  & $7.60\cdot10^{-11}$ & $2.4^{+2.2}_{-1.8}\cdot10^{38}$  & $4.2^{+4.6}_{-3.2}\cdot10^{37}$  \\
IC5052        & 6.0  & 157.3  & $7.66\cdot10^{-2}$ & $1.03\cdot10^{9}$  & $7.40\cdot10^{-11}$ & $5.4^{+2.0}_{-1.6}\cdot10^{39}$  & $9.6^{+5.7}_{-3.2}\cdot10^{38}$  \\
VV747$^{\ast}$         & 9.2  & 34.8   & $6.40\cdot10^{-4}$ & $8.88\cdot10^{6}$  & $7.20\cdot10^{-11}$ & $4.7^{+2.2}_{-1.7}\cdot10^{39}$  & $8.2^{+5.8}_{-3.2}\cdot10^{38}$  \\
NGC2188       & 7.4  & 14.2   & $6.91\cdot10^{-2}$ & $1.05\cdot10^{9}$  & $6.60\cdot10^{-11}$ & $5.7^{+5.2}_{-4.1}\cdot10^{38}$  & $1.0^{+1.1}_{-0.7}\cdot10^{38}$  \\
UGC04787      & 20.3 & $-0.9$ & $6.79\cdot10^{-3}$ & $1.08\cdot10^{8}$ & $6.30\cdot10^{-11}$ & $<1.6\cdot10^{39}$              & $<3.0\cdot10^{38}$               \\
NGC3351       & 10.0 & 36.2   & $1.18\cdot10^{0}$  & $1.91\cdot10^{10}$ & $6.20\cdot10^{-11}$ & $5.9^{+3.8}_{-2.9}\cdot10^{39}$  & $1.0^{+0.9}_{-0.5}\cdot10^{39}$  \\
SexA          & 1.3  & 8.3    & $6.57\cdot10^{-3}$ & $1.09\cdot10^{8}$  & $6.00\cdot10^{-11}$ & $<5.1\cdot10^{37}$               & $<9.5\cdot10^{36}$               \\
NGC7090       & 6.7  & 36.4   & $2.58\cdot10^{-1}$ & $4.70\cdot10^{9}$  & $5.50\cdot10^{-11}$ & $2.2^{+1.4}_{-1.1}\cdot10^{39}$  & $3.8^{+3.2}_{-2.0}\cdot10^{38}$  \\
NGC0055       & 2.1  & 674.4  & $1.21\cdot10^{-1}$ & $2.35\cdot10^{9}$  & $5.10\cdot10^{-11}$ & $2.6^{+0.9}_{-0.7}\cdot10^{39}$  & $4.5^{+2.6}_{-1.4}\cdot10^{38}$  \\
\hline
\end{tabular}
}
\tablefoot{Columns list the galaxy name, distance, net counts, star formation rate, stellar mass, specific star formation rate, total luminosity, and (\texttt{apec}) soft-band luminosity of the thermal component. For non-detections, 90\,\% upper limits are given. $^{\ast}$Not included in $L_\mathrm{X}-\mathrm{SFR}$ fit due to unreliable SFR.}
\end{table}

\begin{table}[htbp!]
\caption{Properties of low sSFR galaxy sample.}
\label{tab_low1}
\centering
\footnotesize
\resizebox{\textwidth}{!}{%
\begin{tabular}{lccccccc}
\hline\hline
Name & Dist. & Cts & SFR & $M_\ast$ & sSFR & $L_{\mathrm{X,tot}}^{\mathrm{0.5-8\,keV}}$ & $L_{\mathrm{X,th}}^{\mathrm{0.5-2\,keV}}$ \\
 & (Mpc) &  & ($\mathrm{M_\odot\,yr}^{-1}$) & ($\mathrm{M_\odot}$) & ($\mathrm{yr}^{-1}$) & (\si{erg.s^{-1}}) & (\si{erg.s^{-1}}) \\
\hline
NGC4080        & 15.0 & 3.1    & $5.56\cdot10^{-2}$ & $1.19\cdot10^{9}$  & $4.68\cdot10^{-11}$ & $<1.2\cdot10^{39}$              & $<2.0\cdot10^{38}$              \\
KUG-1157+315   & 23.2 & $-1.0$ & $5.65\cdot10^{-3}$ & $1.26\cdot10^{8}$  & $4.47\cdot10^{-11}$ & $<9.4\cdot10^{38}$              & $<1.7\cdot10^{38}$              \\
DDO088         & 7.7  & 2.1    & $6.88\cdot10^{-3}$ & $1.58\cdot10^{8}$  & $4.37\cdot10^{-11}$ & $<4.2\cdot10^{38}$              & $<6.4\cdot10^{37}$              \\
NGC4517        & 9.7  & 22.0   & $4.35\cdot10^{-1}$ & $1.02\cdot10^{10}$ & $4.27\cdot10^{-11}$ & $1.3^{+1.2}_{-1.0}\cdot10^{39}$ & $<4.2\cdot10^{38}$              \\
IC2171         & 9.9  & 3.4    & $2.39\cdot10^{-2}$ & $6.00\cdot10^{8}$  & $3.98\cdot10^{-11}$ & $<4.7\cdot10^{38}$              & $<8.2\cdot10^{37}$              \\
HIZSS008       & 7.4  & $-1.2$ & $1.93\cdot10^{-2}$ & $4.91\cdot10^{8}$  & $3.93\cdot10^{-11}$ & $<1.7\cdot10^{37}$              & $<1.3\cdot10^{36}$              \\
AGC200499      & 10.4 & 1.3    & $2.00\cdot10^{-2}$ & $5.25\cdot10^{8}$  & $3.80\cdot10^{-11}$ & $<4.3\cdot10^{38}$              & $<6.8\cdot10^{37}$              \\
ESO269-058     & 3.8  & 4.8    & $1.28\cdot10^{-2}$ & $3.49\cdot10^{8}$  & $3.68\cdot10^{-11}$ & $<1.0\cdot10^{38}$              & $<1.6\cdot10^{37}$              \\
KKH58$^{\ast}$          & 18.4 & $-0.1$ & $6.60\cdot10^{-4}$ & $1.89\cdot10^{7}$  & $3.52\cdot10^{-11}$ & $<4.8\cdot10^{38}$              & $<6.0\cdot10^{37}$              \\
ESO379-024     & 4.9  & $-0.4$ & $2.70\cdot10^{-4}$ & $7.82\cdot10^{6}$  & $3.51\cdot10^{-11}$ & $<1.3\cdot10^{37}$              & $<2.1\cdot10^{36}$              \\

\end{tabular}
}
\end{table}

\begin{table}[htbp!]
\label{tab_low2}
\centering
\footnotesize
\resizebox{\textwidth}{!}{%
\begin{tabular}{lccccccc}
UGC07131       & 16.8 & 1.0    & $2.20\cdot10^{-4}$ & $6.29\cdot10^{6}$  & $3.44\cdot10^{-11}$ & $<9.1\cdot10^{38}$              & $<1.3\cdot10^{38}$              \\
CGCG-095-078$^{\ast}$   & 7.4  & 0.4    & $7.30\cdot10^{-4}$ & $2.15\cdot10^{7}$  & $3.42\cdot10^{-11}$ & $<1.2\cdot10^{38}$              & $<1.6\cdot10^{37}$              \\
KKSG9          & 10.0 & $-0.8$ & $4.11\cdot10^{-3}$ & $1.28\cdot10^{8}$  & $3.22\cdot10^{-11}$ & $<9.7\cdot10^{37}$              & $<1.7\cdot10^{37}$              \\
CGMW1-260      & 10.8 & 3.3    & $6.57\cdot10^{-2}$ & $2.17\cdot10^{9}$  & $3.02\cdot10^{-11}$ & $<7.3\cdot10^{38}$              & $<1.2\cdot10^{38}$              \\
IC2684         & 10.3 & 3.0    & $4.90\cdot10^{-4}$ & $1.65\cdot10^{7}$  & $2.98\cdot10^{-11}$ & $<7.4\cdot10^{38}$              & $<1.2\cdot10^{38}$              \\
IC5201         & 8.8  & $-2.9$ & $2.73\cdot10^{-1}$ & $9.46\cdot10^{9}$  & $2.88\cdot10^{-11}$ & $<6.7\cdot10^{38}$              & $<8.1\cdot10^{37}$              \\
NGC7462        & 10.1 & 4.7    & $1.50\cdot10^{-1}$ & $5.32\cdot10^{9}$  & $2.82\cdot10^{-11}$ & $<9.2\cdot10^{38}$              & $<1.5\cdot10^{38}$              \\
MCG\mbox{--}03-34-002 & 7.9 & 41.3 & $3.13\cdot10^{-3}$ & $1.14\cdot10^{8}$ & $2.75\cdot10^{-11}$ & $1.6^{+0.7}_{-0.6}\cdot10^{39}$ & $<4.6\cdot10^{38}$              \\
ESO137-018     & 6.4  & 142.6  & $4.49\cdot10^{-2}$ & $1.67\cdot10^{9}$  & $2.69\cdot10^{-11}$ & $3.2^{+1.2}_{-0.9}\cdot10^{39}$ & $<8.9\cdot10^{38}$              \\
VCC1675        & 7.5  & 1.6    & $3.02\cdot10^{-3}$ & $1.23\cdot10^{8}$  & $2.45\cdot10^{-11}$ & $<2.2\cdot10^{38}$              & $<3.5\cdot10^{37}$              \\
ESO501-023     & 7.0  & 181.4  & $1.47\cdot10^{-2}$ & $6.12\cdot10^{8}$  & $2.40\cdot10^{-11}$ & $6.9^{+2.5}_{-1.8}\cdot10^{39}$ & $<1.9\cdot10^{39}$              \\
UGC7321        & 17.2 & 0.9    & $8.68\cdot10^{-2}$ & $3.97\cdot10^{9}$  & $2.19\cdot10^{-11}$ & $<1.2\cdot10^{39}$              & $<1.7\cdot10^{38}$              \\
UGC05672       & 11.4 & 1.0    & $5.75\cdot10^{-3}$ & $2.75\cdot10^{8}$  & $2.09\cdot10^{-11}$ & $<6.8\cdot10^{38}$              & $<9.8\cdot10^{37}$              \\
ESO376-016     & 7.1  & 2.6    & $1.50\cdot10^{-4}$ & $7.21\cdot10^{6}$  & $2.06\cdot10^{-11}$ & $<1.9\cdot10^{38}$              & $<3.1\cdot10^{37}$              \\
ESO558-011     & 8.4  & 4.3    & $7.13\cdot10^{-3}$ & $3.49\cdot10^{8}$  & $2.04\cdot10^{-11}$ & $2.1^{+2.7}_{-1.9}\cdot10^{38}$ & $<7.2\cdot10^{37}$              \\
CGCG-035-007   & 13.0 & 2.1    & $5.40\cdot10^{-4}$ & $2.67\cdot10^{7}$  & $2.01\cdot10^{-11}$ & $<8.7\cdot10^{38}$              & $<1.4\cdot10^{38}$              \\
UGC05797       & 7.0  & $-2.6$ & $1.12\cdot10^{-3}$ & $6.15\cdot10^{7}$  & $1.82\cdot10^{-11}$ & $<9.1\cdot10^{37}$              & $<1.4\cdot10^{37}$              \\
KDG082         & 16.6 & 0.7    & $4.65\cdot10^{-3}$ & $2.61\cdot10^{8}$  & $1.78\cdot10^{-11}$ & $<1.1\cdot10^{39}$              & $<1.5\cdot10^{38}$              \\
VCC2033        & 9.8  & 5.4    & $1.04\cdot10^{-3}$ & $5.94\cdot10^{7}$  & $1.75\cdot10^{-11}$ & $4.0^{+4.0}_{-2.8}\cdot10^{38}$ & $<1.3\cdot10^{38}$              \\
KKSG38         & 10.4 & $-0.5$ & $3.30\cdot10^{-4}$ & $1.95\cdot10^{7}$  & $1.70\cdot10^{-11}$ & $<9.4\cdot10^{37}$              & $<1.1\cdot10^{37}$              \\
UGC05923       & 22.3 & 0.4    & $1.23\cdot10^{-2}$ & $8.10\cdot10^{8}$  & $1.51\cdot10^{-11}$ & $<1.1\cdot10^{39}$              & $<1.5\cdot10^{38}$              \\
IC3247         & 24.4 & $-3.2$ & $1.85\cdot10^{-2}$ & $1.19\cdot10^{9}$  & $1.55\cdot10^{-11}$ & $<7.6\cdot10^{38}$              & $<1.2\cdot10^{38}$              \\
UGC07596       & 4.6  & 0.2    & $5.50\cdot10^{-4}$ & $3.55\cdot10^{7}$  & $1.55\cdot10^{-11}$ & $<4.2\cdot10^{37}$              & $<5.6\cdot10^{36}$              \\
UGCA365        & 5.2  & $-0.8$ & $6.30\cdot10^{-4}$ & $4.18\cdot10^{7}$  & $1.51\cdot10^{-11}$ & $<3.1\cdot10^{37}$              & $<3.7\cdot10^{36}$              \\
PGC170257      & 6.5  & 6.5    & $7.50\cdot10^{-4}$ & $5.06\cdot10^{7}$  & $1.49\cdot10^{-11}$ & $1.6^{+1.7}_{-1.2}\cdot10^{38}$ & $<5.3\cdot10^{37}$              \\
CGMW2-3473     & 7.4  & $-0.8$ & $1.59\cdot10^{-3}$ & $1.13\cdot10^{8}$  & $1.41\cdot10^{-11}$ & $<6.1\cdot10^{37}$              & $<1.1\cdot10^{37}$              \\
ESO496-010     & 7.8  & $-1.0$ & $1.18\cdot10^{-3}$ & $8.49\cdot10^{7}$  & $1.39\cdot10^{-11}$ & $<9.5\cdot10^{37}$              & $<1.7\cdot10^{37}$              \\
MCG-02-33-075  & 8.7  & $-7.0$ & $4.49\cdot10^{-3}$ & $3.49\cdot10^{8}$  & $1.29\cdot10^{-11}$ & $<1.6\cdot10^{38}$              & $<2.9\cdot10^{37}$              \\
UGC05288       & 6.8  & 1.6    & $1.48\cdot10^{-3}$ & $1.12\cdot10^{8}$  & $1.32\cdot10^{-11}$ & $<2.2\cdot10^{38}$              & $<3.5\cdot10^{37}$              \\
ESO272-025     & 5.9  & $-4.0$ & $7.10\cdot10^{-4}$ & $5.83\cdot10^{7}$  & $1.21\cdot10^{-11}$ & $<4.6\cdot10^{37}$              & $<7.1\cdot10^{36}$              \\
MAPS1206+31    & 7.4  & $-1.0$ & $3.60\cdot10^{-4}$ & $3.16\cdot10^{7}$  & $1.14\cdot10^{-11}$ & $<9.0\cdot10^{37}$              & $<1.7\cdot10^{37}$              \\
NGC3368        & 10.4 & 70.1   & $5.26\cdot10^{-1}$ & $4.79\cdot10^{10}$ & $1.10\cdot10^{-11}$ & $7.1^{+3.5}_{-2.6}\cdot10^{39}$ & $<2.0\cdot10^{39}$              \\
KUG-1202+286   & 6.7  & 0.7    & $4.30\cdot10^{-4}$ & $4.05\cdot10^{7}$  & $1.05\cdot10^{-11}$ & $<1.0\cdot10^{38}$              & $<1.5\cdot10^{37}$              \\
ESO269-037     & 3.5  & $-0.1$ & $2.00\cdot10^{-5}$ & $2.53\cdot10^{6}$  & $9.87\cdot10^{-12}$ & $<7.8\cdot10^{36}$              & $<1.0\cdot10^{36}$              \\
IC0559$^{\ast}$         & 9.4  & 0.3    & $3.00\cdot10^{-4}$ & $3.29\cdot10^{7}$  & $9.12\cdot10^{-12}$ & $<2.3\cdot10^{38}$              & $<3.2\cdot10^{37}$              \\
Phoenix        & 0.4  & 1.8    & $1.00\cdot10^{-5}$ & $9.07\cdot10^{5}$  & $8.91\cdot10^{-12}$ & $<1.3\cdot10^{36}$              & $<1.8\cdot10^{35}$              \\
ESO274-001     & 3.1  & 15.8   & $1.78\cdot10^{-2}$ & $2.18\cdot10^{9}$  & $8.13\cdot10^{-12}$ & $<1.5\cdot10^{38}$              & $<2.5\cdot10^{37}$              \\
IC3077         & 9.1  & 1.0    & $1.79\cdot10^{-3}$ & $2.25\cdot10^{8}$  & $7.94\cdot10^{-12}$ & $<2.8\cdot10^{38}$              & $<4.1\cdot10^{37}$              \\
NGC5398        & 8.1  & 82.7   & $9.17\cdot10^{-3}$ & $1.34\cdot10^{9}$  & $6.83\cdot10^{-12}$ & $2.9^{+1.2}_{-0.9}\cdot10^{39}$ & $<8.3\cdot10^{38}$              \\
IC4710         & 7.9  & $-0.5$ & $8.26\cdot10^{-3}$ & $1.25\cdot10^{9}$  & $6.61\cdot10^{-12}$ & $<3.1\cdot10^{38}$              & $<4.1\cdot10^{37}$              \\
UGC03755       & 7.4  & 4.0    & $4.00\cdot10^{-4}$ & $6.61\cdot10^{7}$  & $6.00\cdot10^{-12}$ & $<4.4\cdot10^{38}$              & $<7.5\cdot10^{37}$              \\
ESO270-017     & 3.6  & $-6.6$ & $5.76\cdot10^{-2}$ & $9.78\cdot10^{9}$  & $5.89\cdot10^{-12}$ & $<4.4\cdot10^{37}$              & $<4.7\cdot10^{36}$              \\
IC3718         & 6.8  & $-1.8$ & $1.16\cdot10^{-3}$ & $2.31\cdot10^{8}$  & $5.01\cdot10^{-12}$ & $<4.3\cdot10^{37}$              & $<4.2\cdot10^{36}$              \\
IC3104         & 2.3  & 8.1    & $6.00\cdot10^{-5}$ & $2.01\cdot10^{7}$  & $2.94\cdot10^{-12}$ & $<2.6\cdot10^{37}$              & $<4.3\cdot10^{36}$              \\
ESO497-004     & 19.0 & $-0.3$ & $1.00\cdot10^{-4}$ & $4.04\cdot10^{7}$  & $2.48\cdot10^{-12}$ & $<2.2\cdot10^{38}$              & $<3.6\cdot10^{37}$              \\
PGC013294      & 8.0  & $-2.0$ & $2.00\cdot10^{-5}$ & $8.82\cdot10^{6}$  & $1.90\cdot10^{-12}$ & $<3.5\cdot10^{37}$              & $<4.9\cdot10^{36}$              \\
IC2038$^{\ast}$         & 19.2 & 30.9   & $1.00\cdot10^{-5}$ & $1.03\cdot10^{7}$  & $8.14\cdot10^{-13}$ & $2.0^{+1.0}_{-0.8}\cdot10^{39}$ & $<5.7\cdot10^{38}$              \\
\hline
\end{tabular}
}
\tablefoot{Columns list the galaxy name, distance, net counts, star formation rate, stellar mass, specific star formation rate, total luminosity, and (\texttt{apec}) soft-band luminosity of the thermal component. For non-detections, 90\,\% upper limits are given. $^{\ast}$Not included in $L_\mathrm{X}-\mathrm{SFR}$ fit due to unreliable SFR.}
\end{table}

\begin{table}[htbp!]
\caption{Properties of early-type galaxy sample.}
\label{tab_etg}
\centering
\footnotesize
\resizebox{\textwidth}{!}{%
\begin{tabular}{lccccccc}
\hline\hline
Name & Dist. & Cts & SFR & $M_\ast$ & sSFR & $L_{\mathrm{X,tot}}^{\mathrm{0.5-8\,keV}}$ & $L_{\mathrm{X,th}}^{\mathrm{0.5-2\,keV}}$ \\
 & (Mpc) &  & ($\mathrm{M_\odot\,yr}^{-1}$) & ($\mathrm{M_\odot}$) & ($\mathrm{yr}^{-1}$) & (\si{erg.s^{-1}}) & (\si{erg.s^{-1}}) \\
\hline
ESO294-010  & 1.9  & 0.8    & $2.00\cdot10^{-5}$ & $2.14\cdot10^{6}$  & $1.16\cdot10^{-11}$ & $<1.2\cdot10^{37}$              & $<1.8\cdot10^{36}$ \\
ESO384-016  & 4.5  & 0.4    & $1.00\cdot10^{-4}$ & $1.13\cdot10^{7}$  & $8.74\cdot10^{-12}$ & $<4.6\cdot10^{37}$              & $<6.4\cdot10^{36}$ \\
NGC5237     & 3.4  & $-3.5$ & $5.70\cdot10^{-4}$ & $1.30\cdot10^{8}$  & $4.39\cdot10^{-12}$ & $<1.1\cdot10^{37}$              & $<9.3\cdot10^{35}$ \\
NGC1400     & 24.5 & 80.5   & $2.05\cdot10^{-1}$ & $4.70\cdot10^{10}$ & $4.37\cdot10^{-12}$ & $2.2^{+1.0}_{-0.7}\cdot10^{40}$ & $<6.3\cdot10^{39}$ \\
NGC1291     & 8.8  & 411.8  & $1.60\cdot10^{-1}$ & $4.95\cdot10^{10}$ & $3.24\cdot10^{-12}$ & $7.9^{+2.9}_{-2.1}\cdot10^{39}$ & $<2.2\cdot10^{39}$ \\
NGC3115     & 9.7  & 61.9   & $1.42\cdot10^{-1}$ & $5.79\cdot10^{10}$ & $2.46\cdot10^{-12}$ & $5.4^{+2.5}_{-1.9}\cdot10^{39}$ & $<1.5\cdot10^{39}$ \\
NGC4308     & 14.0 & 1.0    & $1.88\cdot10^{-3}$ & $8.22\cdot10^{8}$  & $2.29\cdot10^{-12}$ & $<6.0\cdot10^{38}$              & $<8.9\cdot10^{37}$ \\
NGC5102     & 3.4  & 52.3   & $5.94\cdot10^{-3}$ & $2.65\cdot10^{9}$  & $2.24\cdot10^{-12}$ & $2.8^{+2.3}_{-1.8}\cdot10^{38}$ & $<8.6\cdot10^{37}$ \\
ESO269-066  & 3.8  & 8.8    & $1.00\cdot10^{-5}$ & $9.47\cdot10^{6}$  & $1.23\cdot10^{-12}$ & $<1.2\cdot10^{38}$              & $<2.1\cdot10^{37}$ \\
NGC3384     & 11.4 & 9.7    & $1.74\cdot10^{-2}$ & $2.40\cdot10^{10}$ & $7.24\cdot10^{-13}$ & $<2.6\cdot10^{39}$              & $<4.6\cdot10^{38}$ \\
NGC6684     & 8.7  & 26.0   & $1.71\cdot10^{-2}$ & $2.64\cdot10^{10}$ & $6.46\cdot10^{-13}$ & $1.3^{+0.9}_{-0.7}\cdot10^{39}$ & $<3.8\cdot10^{38}$ \\
NGC3412     & 10.4 & 13.0   & $4.13\cdot10^{-3}$ & $1.19\cdot10^{10}$ & $3.47\cdot10^{-13}$ & $1.6^{+1.3}_{-1.0}\cdot10^{39}$ & $<4.9\cdot10^{38}$ \\
NGC3377     & 10.9 & 25.0   & $9.00\cdot10^{-4}$ & $1.42\cdot10^{10}$ & $6.30\cdot10^{-14}$ & $3.2^{+2.1}_{-1.6}\cdot10^{39}$ & $<9.5\cdot10^{38}$ \\
\hline
\end{tabular}
}
\tablefoot{Columns list the galaxy name, distance, net counts, star formation rate, stellar mass, specific star formation rate, total luminosity, and (\texttt{apec}) soft-band luminosity of the thermal component. For non-detections, 90\,\% upper limits are given.}
\end{table}

\end{appendix}

\end{document}